\documentclass{article}
\pdfoutput=1
\usepackage{geometry}
\usepackage[utf8]{inputenc}
\usepackage[T1]{fontenc}
\usepackage{lmodern}

\usepackage[dvipsnames]{xcolor}

\usepackage{tikz}
\usetikzlibrary{positioning}
\usetikzlibrary{patterns}

\usepackage{amsmath,amssymb,amsthm}
\usepackage[sort&compress,numbers]{natbib}

\usepackage{hyperref}

\usepackage{color}
\newcommand{\tc}[2]{\textcolor{#1}{#2}}
\definecolor{red}{rgb}{1,0,0} 

\definecolor{grey}{rgb}{.5,.5,.5} 

 \definecolor{darkgreen}{rgb}{.3, .7, .3}

\newcommand{\green}[1]{\tc{darkgreen}{#1}}
 \definecolor{purple}{rgb}{.7, 0, 1}

\newcommand{\td}[1]{\widetilde{#1}}

\newcommand{\R}{\mathbb{R}}
\newcommand{\C}{\mathbb{C}}
\newcommand{\dd}{\textrm{d}}

\newcommand{\bigO}{\mathcal{O}}

\newtheorem{theorem}{Theorem}

\theoremstyle{definition}

\newtheorem{conjecture}[theorem]{Conjecture}

\theoremstyle{remark}

\hypersetup{pdfauthor={Michael Borinsky, David Broadhurst},
            pdftitle={Resonant resurgent asymptotics from quantum field theory},
            pdfsubject={},
            pdfkeywords={Renormalon, Resurgence, trans-series, Borel transform, non-perturbative, Quantum Field Theory, Renormalization, Hopf algebra, Scalar Phi3 Theory, Asymptotics, Differential equation, ODE}
  linkcolor  = RedViolet!85!black,
  citecolor  = BlueGreen!85!black,
  urlcolor   = YellowOrange!85!black,
  colorlinks = true,
}

\begin{document}
\title{Resonant resurgent asymptotics from quantum field theory}
\author{
Michael Borinsky\\[1ex]
\small{Institute for Theoretical Studies}\\
\small{ETH Z\"urich}\\
\small{8092 Z\"urich, Switzerland}\\
 \and
David Broadhurst\\[1ex]
\small{School of Physical Sciences}\\
\small{Open University}\\
\small{Milton Keynes MK7 6AA, UK}
}
\date{%
}

\maketitle
\begin{abstract}
We perform an all-order resurgence analysis of a quantum field theory renormalon that contributes to
an anomalous dimension in six-dimensional scalar $\phi^3$ theory and is governed
by a third-order nonlinear differential equation.
We augment the factorially divergent perturbative expansion associated to the renormalon
by asymptotic expansions to all instanton orders, in a conjectured and well-tested formula.
A distinctive feature of this renormalon singularity is the appearance of logarithmic terms,
starting at second-instanton order in the trans-series.
To highlight this and to illustrate our methods, we also analyze
the trans-series for a closely related second-order nonlinear differential equation that
exhibits a similarly resonant structure but lacks logarithmic contributions.
\end{abstract}

\section{Introduction}
\subsection{Renormalons and resurgence}

Perturbative computations in quantum field theories require the treatment of various kinds of divergences. For instance, renormalization is needed at each order in the perturbative expansion to bring UV-singularities under control. By now, the physical and mathematical reasoning behind this procedure is very well understood. 
A much less explored source of concern is the divergent nature \emph{of the perturbative expansion itself}~\cite{Dyson:1952tj}. 
In this article we shall illuminate some of the technicalities which are necessary to make sense of such a divergent expansion via \emph{resurgence}.

Roughly, we can distinguish two types of sources for a divergence of the perturbative expansion: a non-perturbative contribution that stems from a nontrivial solution of the classical field equations, which is referred to as an \emph{instanton solution}. This type of divergence can also be observed in quantum mechanical models~\cite{Lam:1968tk,Bender:1973rz,Lipatov:1976ny,Zinn-Justin:2004vcw,Zinn-Justin:2004qzw} and entirely combinatorial or topological models~\cite{Borinsky:2017hkb,borinsky2020euler} (see also~\cite{Zinn-Justin:1980oco,le2012large,marino2015instantons} for exhaustive introductions into such semi-classical instanton phenomena and calculations).

A further type of divergence is generated by renormalization subtraction terms of a certain set of Feynman diagrams. Such a divergence is called \emph{a renormalon}~\cite{tHooft:1977xjm,Lautrup:1977hs,Parisi:1978bj,Beneke:1998ui,Shifman:2013uka} and can often be explicitly associated with a specific large-$N$ limit of the underlying theory~\cite{Palanques-Mestre:1983ogz,Broadhurst:1992si,Beneke:1992ea,Beneke:1993yn,Gracey:1996he,Dondi:2020qfj,Dondi:2021buw,DiPietro:2021yxb,Fujimori:2021oqg}.
In this article we will deal with a specific UV-renormalon, i.e.\ a renormalon that originates from the UV-subtraction terms of a specific class of diagrams.
Recently, Mari\~no and Reis showed that also super-renormalizable QFTs and integrable models can contain infrared-renormalon singularities~\cite{Marino:2019eym,Marino:2019fvu} while making use of technology from~\cite{Broadhurst:1992si} (see also~\cite{Marino:2021dzn}).  A direct approach to renormalon singularities based on renormalization group considerations has also been recently put forward~\cite{Maiezza:2018pkk,Antipin:2018asc,Maiezza:2020nbe,Maiezza:2021mry,Caprini:2020lff}. Moreover, it has been shown recently that also quantum mechanical models can feature renormalon singularities~\cite{Pazarbasi:2019web}.
Another recent development is the observation that the usually applied resurgence framework which works with \emph{Gevrey-1} sequences might have to be generalized to be able to handle certain   QFT renormalon singularities~\cite{Cavalcanti:2020osb}. In this work, we will only deal with \emph{Gevrey-1} singularities. 
The role of the renormalization scheme in the scope of renormalon divergences was recently studied in~\cite{Balduf:2021kag}. 

\subsection{A renormalon in \texorpdfstring{$\phi^3$}{phi3} theory in six-dimensional spacetime}

We will analyze the resurgence structure of a renormalon that contributes to the perturbative solution of six-dimensional scalar $\phi^3$ theory. This model is especially interesting as it is believed to be asymptotically free~\cite{Macfarlane:1974vp}, its renormalization group functions are known up to a high perturbative order~\cite{Gracey:2015tta,Borinsky:2021jdb,Borinsky:2021gkd,Kompaniets:2021hwg,Borinsky:2022}, the structure of its semi-classical instanton solutions is relatively well-studied~\cite{Brezin:1976vw,Mckane:1978me,Houghton:1978dt,alvarez1988coupling,alvarez1995bender} and, augmented with a bi-adjoint group structure, it is believed to be BCJ dual to Yang--Mills theory~\cite{Bern:2008qj,delaCruz:2017zqr,Arkani-Hamed:2017mur}. Moreover, $\phi^3$ theory in six dimensions has direct application to percolation theory~\cite{Bonfim_1980,Bonfim_1981,Gracey:2015tta,Borinsky:2021jdb} and the Lee--Yang edge singularity~\cite{Fisher_1978,Gracey:2015tta,Borinsky:2021jdb}.

The Feynman diagrams that contribute to this $\phi^3$ theory renormalon are the one-loop self-energy correction together with all possible recursive insertions of this diagram into itself. This set of diagrams can be depicted via the Dyson--Schwinger equation,
\begin{align} \label{dse} \begin{tikzpicture}[baseline={([yshift=1ex]current bounding box.south)}] \coordinate (in); \coordinate[right=.25 of in] (v1); \coordinate[right=.25 of v1] (m); \coordinate[right=.25 of m] (v2); \coordinate[right=.25 of v2] (out); \filldraw[preaction={fill,white},pattern=north east lines] (m) circle(.25); \filldraw (v1) circle(1pt); \filldraw (v2) circle(1pt); \draw (in) -- (v1); \draw (v2) -- (out); \end{tikzpicture} = \frac12 ~ \begin{tikzpicture}[baseline={([yshift=1ex]current bounding box.south)}] \coordinate (in); \coordinate[right=.25 of in] (v1); \coordinate[right=.25 of v1] (vm); \coordinate[right=.25 of vm] (v2); \coordinate[right=.25 of v2] (out); \filldraw (v1) circle(1pt); \filldraw (v2) circle(1pt); \draw (in) -- (v1); \draw (v2) -- (out); \draw (vm) circle (.25); \end{tikzpicture} + \begin{tikzpicture}[baseline={([yshift=1ex]current bounding box.south)}] \coordinate (in); \coordinate[right=.25 of in] (v1); \coordinate[right=.25 of v1] (v2); \coordinate[right=.25 of v2] (m); \coordinate[right=.25 of m] (v3); \coordinate[right=.25 of v3] (v4); \coordinate[right=.25 of v4] (out); \filldraw[preaction={fill,white},pattern=north east lines] (m) circle(.25); \filldraw (v1) circle(1pt); \filldraw (v2) circle(1pt); \filldraw (v3) circle(1pt); \filldraw (v4) circle(1pt); \draw (in) -- (v1); \draw (v1) -- (v2); \draw (v3) -- (v4); \draw (v4) -- (out); \draw (v1) arc (180:0:.5); \end{tikzpicture} + \begin{tikzpicture}[baseline={([yshift=1ex]current bounding box.south)}] \coordinate (in); \coordinate[right=.25 of in] (v1); \coordinate[right=.25 of v1] (v2); \coordinate[right=.25 of v2] (m1); \coordinate[right=.25 of m1] (v3); \coordinate[right=.25 of v3] (v4); \coordinate[right=.25 of v4] (m2); \coordinate[right=.25 of m2] (v5); \coordinate[right=.25 of v5] (v6); \coordinate[right=.25 of v6] (out); \filldraw[preaction={fill,white},pattern=north east lines] (m1) circle(.25); \filldraw[preaction={fill,white},pattern=north east lines] (m2) circle(.25); \filldraw (v1) circle(1pt); \filldraw (v2) circle(1pt); \filldraw (v3) circle(1pt); \filldraw (v4) circle(1pt); \filldraw (v5) circle(1pt); \filldraw (v6) circle(1pt); \draw (in) -- (v1); \draw (v1) -- (v2); \draw (v3) -- (v4); \draw (v5) -- (v6); \draw (v6) -- (out); \draw (v1) arc (180:0:.875); \end{tikzpicture} + \cdots \end{align} 
After renormalization the contribution from this set of diagrams turns out to yield a factorially divergent power series with highly nontrivial properties. 
Using a Hopf-algebraic momentum subtraction renormalization procedure~\cite{Connes:1999yr}, Broadhurst and Kreimer were able to deduce a third-order nonlinear differential equation for the associated contribution to the field anomalous dimension of $\phi^3$ theory in six spacetime dimensions~\cite[Equation~(52)]{BK2}. Remarkably, this equation can be brought into a simpler, factored form (see~\cite[Equation~(53-57)]{BK2} and \cite[Equation~(4.1)]{BDM}). It reads,
\begin{equation}
(g(x)P-1)(g(x)P-2)(g(x)P-3)g(x)=-3,\quad \text{ where }\quad P=x\left(2x\frac{\dd}{{\dd}x}+1\right).\label{ode}
\end{equation}
As a third-order ODE, this equation 
is expected to have a three-dimensional solution space. Surprisingly, due to an irregular singular point at the origin, the equation has \emph{only one unique} formal power series or \emph{perturbative} solution around $x=0$:
\begin{align} g_0(x)=A(x)=\sum_{n=0}^\infty A_nx^n=\tfrac{1}{2}+\tfrac{11}{24}x +\tfrac{47}{36}x^2+\tfrac{2249}{384}x^3+\tfrac{356789}{10368}x^4 +\tfrac{60819625}{248832}x^5+\bigO(x^6). \label{g0} \end{align} 
This formal power series solution was developed to 500 terms by Broadhurst and Kreimer~\cite{BK2}, who observed that the expansion is factorially divergent and therefore constitutes a renormalon contribution to the $\phi^3$ theory field anomalous dimension. The three-dimensional family of other solutions to~(\ref{ode}) is hidden behind exponentially suppressed \emph{non-perturbative} corrections to the solution~(\ref{g0}). 

\subsection{Resurgence}

Resurgence is the mathematical theory of such factorially divergent power series which allows to associate concrete functions to them. It has been developed by \'Ecalle~\cite{ecalle1981fonctions} (see~\cite[Part~II]{Mitschi:2016fxp} for a mathematics focused introduction and~\cite{Marino:2012zq,Dorigoni:2014hea,marino2015instantons,Aniceto:2018bis} for reviews focused on applications in physics). 
Resurgence provides a promising approach to make non-perturbative predictions in quantum field theory 
and string theory 
 (see for instance~\cite{Delabaere1997,delabaere99,alvarez04,Argyres:2012ka,Dunne:2012ae,Dunne:2013ada,Marino:2007te,Pasquetti:2010bps,Aniceto:2011nu,Grassi:2014cla}). Moreover, by its intricate relationship to quantization, resurgence suggest various interesting connections of quantum theory to wall-crossing phenomena, topological recursion, three-manifolds, knot theory, combinatorics and other parts of mathematics~\cite{Marino:2002fk,Andersen:2018khh,Eynard:2019mps,Kontsevich:2020piu,Garoufalidis:2020nut,borinsky2018generating}.

In this article we shall investigate the detailed resurgence properties of the renormalon described by~(\ref{ode}) and its perturbative solution~(\ref{g0}). 
The singular structure of this renormalon features the complete set of divergences, including logarithmic terms, that are expected from a general QFT solution. 
These aspects make this renormalon a particularly interesting and instructive instance of this type of QFT divergence.
Related studies of renormalons based on analysis of Dyson--Schwinger equations have been performed for the Wess--Zumino model~\cite{Bellon:2014zxa,Bellon:2016mje,Clavier:2019sph}, for six-dimensional scalar $\phi^3$ theory in a more general context which also includes vertex corrections~\cite{Bellon:2020uzi,Bellon:2020qlx} and for Yukawa theory~\cite{BD}, where an \emph{all-order} trans-series solution of the Dyson--Schwinger equation could be achieved. 

\subsection{Resonant resurgence phenomena and logarithmic terms}

In~\cite{BK1,BK2}, the expansion~(\ref{g0}) occurs at $x<0$ and
hence is amenable to Pad\'e--Borel summation. On first sight this is the most relevant case as 
$$\gamma(\lambda) = - \frac{1}{3} \frac{\lambda^2}{(4\pi)^3} \cdot g\left(- \frac{1}{3} \frac{\lambda^2}{(4\pi)^3}\right)$$
is the contribution of~(\ref{g0}) 
to the field anomalous dimension of 
six-dimensional $\phi^3$-theory with real coupling constant 
$\lambda$ in the momentum subtraction scheme.
Recently, the first author, Dunne and Meynig~\cite{BDM} considered
the case with $x>0$ of~(\ref{g0}) which corresponds to the analysis of a QFT with an \emph{imaginary} coupling constant such that $\lambda^2 < 0$.
Physically this is a relevant case as well, as it describes the \emph{Lee--Yang} edge singularity~\cite{Fisher_1978}. 
From a resurgence perspective, the sign choice $x > 0$ is more interesting as 
naive Pad\'e--Borel resummation is not sufficient to make sense of the perturbative expansion.
The series is not Borel-resummable and the resulting ambiguities have to be dealt with explicitly. 
This case will also serve as our entry point into the resurgence analysis of the renormalon described by~(\ref{dse}).

The very first step for such a resurgence analysis of a highly nonlinear ODE is to consider its \emph{linearized} variant. Solving this linearized equation (see~\cite[Eq.~(5.3)]{BDM}) which is homogeneous, one encounters 
three non-perturbative solutions $h_1,h_2,h_3$. These solutions can be obtained by setting 
\begin{equation}
g(x)=g_0(x)+\sigma_k\left(x^{-\frac{35}{12}}\,e^{-\frac{1}{x}}\right)^k h_k(x)+\bigO(\sigma_k^2),\quad k\in \{1,2,3\},\label{hk}
\end{equation}
in~(\ref{ode}) and discarding terms of order $\sigma_k^2$. The parameters 
$\sigma_1,\sigma_2,\sigma_3$ are the integration constants of the linearized homogeneous ODE. Eventually they will parameterize the three-dimensional \emph{non-perturbative} solution space that is 
 expected from a third-order ODE~(\ref{ode}). 

From the first-order contributions in the $\sigma_1,\sigma_2,\sigma_3$ parameters in~(\ref{hk}), one can anticipate a \emph{trans-series} solution of~(\ref{ode}) of the form 
\begin{gather} \label{basictran} g(x)= \sum_{i=0}^\infty \sum_{j=0}^\infty \sum_{k=0}^\infty \sigma_1^i \sigma_2^j \sigma_3^k \left( x^{-\frac{35}{12}}\,e^{-\frac{1}{x}} \right)^{i+2j+3k} G_{i,j,k}(x) \end{gather}
with $G_{0,0,0}(x) = g_0(x)$,
$G_{1,0,0}(x) = h_1(x)$,
$G_{0,1,0}(x) = h_2(x)$ and
$G_{0,0,1}(x) = h_3(x)$.
It was observed in~\cite{BDM} that the expansions $G_{i,j,k}(x)$ start to contain logarithmic terms from the second order in $\sigma_1$ on.
The first occurrence of a logarithmic contribution is found in the $\sigma_1^2$ term at order $x^5$, with
\begin{gather} \label{Gsimple} G_{2,0,0}(x) = -2 + \tfrac{49}{6}x + \tfrac{13235}{1728}x^2 + \ldots + \tfrac{21265}{2304}x^5\log \left( \frac{x}{c} \right) \left(-1+\tfrac{151}{24}x+\ldots\right) \end{gather}
where $c$ is an ambiguous constant, undetermined by the ODE.
We will discuss the character  of this ambiguity in detail in Section~\ref{sec:resonant}.
Key to our analysis of~(\ref{ode}) will be the representation~(\ref{tran}--\ref{nsig}) of its trans-series solution that accounts for the logarithmic terms in a particularly compact way. 

The fact that no logarithmic terms appear in the closely related renormalon in four-dimensional Yukawa theory~\cite{BK1,BK2,BD} is only slightly surprising, as this simpler four-dimensional renormalon is described by a nonlinear first-order ODE. A prominent nonlinear ODE whose solution features logarithmic terms is the Painlev\'e~I equation~\cite{PainI,Pain1a}. This equation is relevant for matrix models 
and string theory~\cite{Marino:2006hs}.  

A plausible explanation of the appearance of these logarithmic terms is an analogy to quantum mechanics where resonances between classical instanton solutions result in logarithmic terms which themselves lie in correspondence to ambiguities of the resummation procedure~\cite{PhysRevA.33.12,Zinn-Justin:1979jnt,delabaere99,alvarez04}. See also~\cite{Pazarbasi:2021ifb} for a recent application of this correspondence. 

The exponents $2$ and $3$ of $e^{-\frac{1}{x}}$ in~(\ref{hk}), which are associated in the truncated trans-series solution~(\ref{hk}) to the solutions $h_2$ and $h_3$ of the linearized ODE, are integer multiples of the exponent $1$ for the $h_1$ solution. This means, there is a \emph{resonance} between the different solutions to the linearized equation, but this resonant ratio does not fully explain the appearance of logarithmic terms and the intuition gained from quantum mechanics unfortunately fails here.
To prove this point and to illustrate the peculiarity of the appearance of logarithmic terms we will analyze the closely related second-order ODE~(\ref{tode}) in Section~\ref{sec:logfree} which features a resonant ratio of trans-series exponential powers between the solutions to its linearized avatar, but is completely free of logarithmic terms.  This second-order ODE will also serve as a suitable warm-up exercise before tackling the full complexity of~(\ref{ode}) in Section~\ref{sec:resonant}. 

\subsection{All-order trans-series analysis}

Our first main result is the following compact representation of the trans-series solution of~(\ref{ode}):
\begin{gather} g(x)=\sum_{m=0}^\infty g_m(x)y^m,\quad y=x^{-\frac{35}{12}}\,e^{-\frac{1}{x}},\label{tran}\\
g_m(x)=\sum_{i=0}^{\lfloor m/2\rfloor}\sum_{j=0}^{\lfloor (m-2i)/3\rfloor} \sigma_1^{m-2i-3j}\widehat{\sigma}_2^i\widehat{\sigma}_3^jx^{5(i+j)}\sum_{n\ge0}a^{(m)}_{i,j}(n)x^n,\label{mtran}\\
\widehat{\sigma}_2=\sigma_2+\tfrac{21265}{2304}\sigma_1^2\log(x),\quad \widehat{\sigma}_3=\sigma_3+\tfrac{21265}{2304}\sigma_1^3\log(x),\label{nsig} \end{gather}
with $\log(x)$ neatly absorbed by~(\ref{nsig}).  
In analogy to instanton expansions of path integrals, we will call the coefficient $g_m(x)$ in front of $y^m$ the $m$-th \emph{instanton}.
The \emph{instanton action} is reflected by the integer $m$ in the exponent of the exponential in $y$ which dictates the magnitude of the exponential suppression of the respective term for $x\rightarrow 0^+$.

The compact absorption of logarithmic terms in~(\ref{nsig}), which involves another interesting constant $\tfrac{21265}{2304}$, provides strong hints for a \emph{cancellation mechanism}. Such a mechanism is observed in quantum mechanical models~\cite{Zinn-Justin:2004vcw,Dunne:2016qix} where it ensures the reality of the trans-series solution in the presence 
of non-perturbative ambiguities as required by physical constraints.

Even after this convenient absorption of logarithms the rational coefficients $a^{(m)}_{i,j}(n)$ are not uniquely determined by the ODE~(\ref{ode}). There remain three intrinsic ambiguities. 
To understand their nature, we can think of Equations~(\ref{tran}--\ref{nsig}) as an \emph{Ansatz} for $g(x)$ that is parameterized by $\sigma_1,\sigma_2,\sigma_3$ and the coefficients $a^{(m)}_{i,j}(n)$ for all $0 \leq i,j,m,n$ with 
$2i + 3j \leq m$. This Ansatz overparameterizes the function $g(x)$ as we can observe by inspection of~(\ref{tran}--\ref{nsig}): For instance, an arbitrary rescaling of $\sigma_1 \rightarrow C_1 \sigma_1$ can be compensated by a corresponding change of $\sigma_2 \rightarrow C_1^2 \sigma_2$, $\sigma_3 \rightarrow C_1^3 \sigma_3$ and $a^{(m)}_{i,j}(n) \rightarrow C_1^{-m}a^{(m)}_{i,j}(n)$. 
The remaining two ambiguities result from the shifts 
$\sigma_2 \rightarrow \sigma_2 + C_2 \sigma_1^2$ and 
$\sigma_3 \rightarrow \sigma_3+ C_3 \sigma_1^3$ that can be compensated by redefinitions of the $a^{(m)}_{i,j}(n)$ coefficients as well.
We proceed to resolve all three ambiguities systematically by fixing specific $a^{(m)}_{i,j}(n)$ coefficients.
The scaling ambiguity is resolved by fixing
\begin{gather} a^{(1)}_{0,0}(0)=-1.\label{fix1} \end{gather}
The choice of a negative sign has the effect that $a^{(1)}_{0,0}(n)$ will eventually be positive for large $n$.
We resolve the remaining, more intricate shift ambiguities by fixing
\begin{gather} \tfrac12a^{(2)}_{0,0}(5)=\tfrac16a^{(3)}_{0,0}(5)=r_1=\tfrac{32642693907919}{36691771392}.\label{fix2} \end{gather}
Any other choice for the coefficients
$a^{(2)}_{0,0}(5)$ and $a^{(3)}_{0,0}(5)$
would have fulfilled the purpose of resolving the overparameterization of $g(x)$ and of producing a unique  %
trans-series solution for the ODE~\eqref{ode}. Our seemingly ad-hoc choices, which involve 
the large fraction $r_1$, are particularly favourable because they result in a simple \emph{asymptotic behaviour} of the coefficients $a^{(m)}_{i,j}(n)$ for large~$n$.

We emphasize that the three ambiguities above are of different nature than the usual three dimensional solution space of 
a third-order ODE.
In our trans-series~(\ref{tran}--\ref{nsig}) of~\eqref{ode} this solution space is parameterized by the trans-series parameters $\sigma_1,\sigma_2$ and $\sigma_3$.
The ODE~\eqref{ode} and the choices~(\ref{fix1}--\ref{fix2}) only fix the values of the $a^{(m)}_{i,j}(n)$ coefficients while leaving 
the trans-series parameters arbitrary.

Our second main result is Conjecture~\ref{conj:three_inst} which completely describes this asymptotic behaviour (including subleading contributions to all orders). As expected from the general theory of resurgence the operation of taking the $n \rightarrow \infty$ asymptotic limit of a set of coefficients $a^{(m)}_{i,j}(n)$ \emph{closes} among 
the sequences $a^{(m)}_{i,j}$, 
in the sense that the coefficients of the asymptotic expansion of  $a^{(m)}_{i,j}(n)$, for large $n$, 
can be expressed using a linear combination of other coefficients $a^{(m')}_{i',j'}(k)$, beginning at small $k$. 
 We explicitly determine these linear combinations and the associated \emph{connection} constants, the \emph{Stokes constants}, numerically.

The formulation of Conjecture~\ref{conj:three_inst} in its compact form was only possible 
with the specific choices made in~(\ref{fix1}--\ref{fix2}). 
Generic or arguably more canonical choices such as, 
$a^{(2)}_{0,0}(5)=a^{(3)}_{0,0}(5)=0$,
would have led to the appearance of additional terms in this asymptotic behaviour, %
which would have involved the large fraction $r_1$ explicitly. %
We will illustrate our reasoning that led us to the specific choices~\eqref{fix2} in detail in Section~\ref{sec:resonant}.

Before stating Conjecture~\ref{conj:three_inst} in the next section, we will introduce a few 
basic notions from the theory of resurgence. 

\section{All-order resurgent asymptotics}

\subsection{Asymptotic notation}

Most power series considered in this article are \emph{factorially divergent}. This means that the coefficients such as $A_n$ from~(\ref{g0}) grow as a shifted $\Gamma$ function modulated by an exponential, 
\begin{equation}
\label{factorial}
\limsup_{n \rightarrow \infty} \left|A_n/ ( \alpha^n \Gamma(n+\beta))\right| = C,
\end{equation}
with some constants $\alpha, \beta \in \R$ where $\alpha \neq 0$.
For this reason, the formal power series $A(x)=\sum_{n =0}^\infty A_n x^n$ has a vanishing radius of convergence. Expressions such as $\sum_{n=0}^\infty A_n x^n$ are therefore supposed to be interpreted formally as objects in the ring of power series $\R[[x]]$. While working with these power series it is convenient to use following notation:
If $A_n,B_n$ and $C_n$ are sequences of real numbers, then the $\bigO$-notation
$A_n = B_n + \bigO(C_n)$
is a shorthand way to write 
$ \limsup_{n \rightarrow \infty} \left| (A_n-B_n)/C_n \right| < \infty. $
The $\bigO$-notation requires us to specify the limit that is taken. 
In this article, asymptotic expansions that involve the integer variable $n$ refer to the limit $n\rightarrow \infty$. 
Taylor expansions of expressions with continuous variables such as $x$ refer to the limit $x \rightarrow 0$. 

If $A_n$ and $B_n$ are sequences of real numbers and $M_{n,k}$ is a family of real number sequences indexed by $k$, then the \emph{asymptotic expansion notation}
$ A_n \sim \sum_{k \geq 0} M_{n,k} B_k $
is shorthand for the \emph{infinite family of $\bigO$-statements},
$ A_n = \sum_{k = 0}^{R-1} M_{n,k} B_k + \bigO(M_{n,R}) \text{ for all } R\geq 0, $
which all shall hold in the $n \rightarrow \infty$ limit.

\subsection{Intuition from the Borel transform}
\label{sec:borel}

The coefficients $A_n$ in~(\ref{g0}) grow factorially. 
Therefore, $g_0(x) = \sum_{n=0}^\infty A_n x^n$ does not converge for any non-zero value of $x$. 
This is a typical phenomenon in perturbative QFT computations, which first has been properly appreciated by Dyson~\cite{Dyson:1952tj}. In general, it is a difficult task to reconstruct a function from a factorially divergent power series and there exist many different approaches and frameworks that aim to tackle this problem (see for instance~\cite{dingle1973asymptotic,ecalle1981fonctions,berry1991hyperasymptotics,costin2008asymptotics,Mitschi:2016fxp}). A pragmatic solution is to analyze the \emph{Borel transform} $\mathcal B[A](z)$ of the factorially 
 divergent power series $A(x)$:
\begin{gather} \label{borel} \mathcal B[A](z) = \sum_{n = 0}^\infty \frac{A_n}{n!} z^n. \end{gather}
If~(\ref{factorial}) holds, then $\mathcal B[A](z)$ is a function of $z$ which is analytic in a non-vanishing domain around the origin. 
For resurgence in its basic form to be applicable, we have to assume that the function $\mathcal B[A](z)$ can be analytically continued 
to the whole complex plane with the exception of a countable number of isolated singular points. After this analytic continuation has been performed, 
an avatar function  
$\widehat{A}(x)$ of the formal power series $A(x)$ can be constructed by applying the Laplace transform which is inverse to the Borel transform,
\begin{gather} \widehat{A}(x) = \int_0^\infty {\dd} z e^{-z} \mathcal B[A](x z). \end{gather}
In~\cite{BK1}, the case with $x<0$ case was considered, 
in which there are no singularities on the line of integration and which is therefore Borel-resummable. Here, we wish to go beyond this simple situation and consider the $x > 0$ case where the line of integration meets singularities of $\mathcal B[A](z)$.
Due to these singularities, it is a nontrivial problem to evaluate the Laplace transform of $\mathcal B[A](z)$. This problem is solved by Laplace--Borel--\'Ecalle resummation (see~\cite[Ch.~5]{Mitschi:2016fxp} for the details of this procedure). 
Resolving the ambiguities of the resummation in a situation with multiple trans-series parameters is known to require the full resurgence machinery and knowledge of all \emph{Stokes constants}~\cite{Aniceto:2013fka}.
Here, we shall focus on a specific highly nontrivial step in this procedure: the analytic continuation of the function $\mathcal B[A](z)$.

We will assume that the function $\mathcal B[A](z)$ only has singularities on the real axis at the evenly spaced locations $z \in \{1,2,\ldots\}$. This situation is typical for quantum field theory applications and does not imply a serious restriction of the applicability of our methods. Under this assumption, the power series representation~(\ref{borel}) provides an analytic expression for $\mathcal B[A](z)$ which is valid inside the unit disc $\{ z \in \C : |z| < 1\}$. Convergence in a larger domain is obstructed by the first singularity of $\mathcal B[A](z)$ at $z=1$. For the Borel transforms of the power series under consideration in this article, the local expansions of $\mathcal B[A](z)$ in the vicinity of its singularities are of the form,
\begin{gather} \label{borelsing} \mathcal B[A](z) \sim \sum_{m \geq 0} \left(1-\frac{z}{k}\right)^{m-\beta_k} c_{k,m} \quad \text{ as } \quad z \rightarrow k^{-}, \end{gather}
where $\beta_k$ are non-integer rational numbers and $c_{k,m}$ arbitrary coefficients.
By Darboux's theorem the large-order behaviour of the expansion coefficients $\frac{A_n}{n!}$ in~(\ref{borel}) is governed by the singular behaviour of $\mathcal B[A](z)$ for $z\rightarrow 1$. To sketch the complete argument that establishes this relationship, we can assume that $z$ is close to $1$. In this case, we can match both expansions,
\begin{gather} \mathcal B[A](z) = \sum_{n =0}^\infty \frac{A_n}{n!} z^n \sim \sum_{m \geq 0} (1-z)^{m-\beta_1} c_{1,m} \text{ as } z \rightarrow 1^{-}, \end{gather}
and compare coefficients for $n\rightarrow \infty$,
\begin{gather} \frac{A_n}{n!} \sim \sum_{m \geq 0} \binom{n - m + \beta_1 -1}{n} c_{1,m} \text{ as } n \rightarrow \infty. \end{gather}
It follows that
\begin{gather} A_n \sim \sum_{m \geq 0} \Gamma(n-m+\beta_1)\frac{c_{1,m}}{\Gamma(-m+\beta_1)}. \end{gather}
Singular cases can arise if $\beta_1$ is an integer and the $\Gamma$ function on the right hand side can develop a pole. 
This would lead to logarithmic terms in the expansions but does not apply to our cases, where all $\beta_k$ are non-integer rational numbers. 

From this argument, it is evident that the factorially 
 divergent large-order behaviour of the $A_n$ coefficients can be translated into the singular behaviour of the Borel transform at the first singularity and vice versa. The power series $\sum_{m \geq 0} (1-z)^{m-\beta_1} c_{1,m}$ provides an expansion of the function $\mathcal B[A](z)$ which converges in the unit disc around the point $z=1$. A larger radius of convergence is obstructed by the singularity of $\mathcal B[A](z)$ at $z=2$. The asymptotic behaviour of the $c_{1,m}$ coefficients for $m \rightarrow \infty$ is in one-to-one correspondence with the singular expansion around the $z=2$ singularity in~(\ref{borelsing}) by the same reasoning as above. Repeating this process for all further singularities along the real line,  we can perform the analytic continuation procedure by iteratively computing the large-order expansion of the previous expansion.

We will apply this procedure to the power series solution~(\ref{g0}) of the differential equation~(\ref{ode}). A practically useful observation is that not only the explicit values of the initial coefficients are determined by the differential equation, but also information on the various higher-order expansions of the coefficients can be deduced. In fact, all the coefficients of the higher-order expansions of the initial sequence can be determined from the differential equation except for certain overall normalization constants. These normalization constants, the \emph{Stokes constants}, provide the explicit connection coefficients between low-order and large-order behaviour of the various sequences. 

An interesting phenomenon arises at the expansion of the Borel transform $\mathcal B[A](z)$ around the second singularity at $z=2$. The large-order behaviour of the coefficients is determined by two singularities of $\mathcal B[A](z)$ in the complex plane: the one at $z=1$ and the one at $z=3$. The large-order behaviour of the coefficients around the point $z=2$ will therefore be governed by two contributions: one alternating (backwards looking from $2$ to $1$) and one non-alternating (forwards looking from $2$ to $3$). We will observe and discuss this phenomenon in our detailed analysis in Sections~\ref{sec:logfree} and \ref{sec:resonant}.

\subsection{All-order resurgent asymptotics for a resonant renormalon}
Instead of working with the Borel transform of factorially divergent power series directly, it is usually more convenient to keep working with the initial factorially 
divergent power series and augment it by exponentially suppressed \emph{trans-series} contributions as we did in~(\ref{tran}--\ref{nsig}). A higher-order trans-series term of the form $x^{-\beta_k} e^{-k/x}$ can directly be associated to a expansion of the Borel transform at the singular point $z=k$. 

For our concrete trans-series solution~(\ref{tran}) of the ODE~(\ref{ode}) this means that the $m$-th instanton coefficient $g_m(x)$ in front of $\left(x^{-\frac{35}{12}} e^{-\frac{1}{x}}\right)^m$ encodes the local expansion around the $m$-th pole of the Borel transform of the perturbative solution $g_0(x)= A(x)$.

 An advantage of this trans-series based approach is that the low- and large-order correspondence works transparently with a minimal number of transcendental prefactors and that the \emph{trans-series Ansatz}, which is substituted into the ODE together with a suitable choices for all ambiguities, yields a mechanical way to generate the respective higher-order expansions.
Moreover, the trans-series approach is compatible with \emph{alien calculus}, which constitutes a corner stone of resurgence theory. We will illustrate this compatibility in the scope of the analytic continuation process in Section~\ref{sec:alien}.

With these mathematical tools at hand we are able to formulate our second main result: A conjectured though well-tested solution of the analytic continuation problem for ODE~(\ref{ode}) at \emph{all orders}. This result has been obtained by employing a mixture of empirical and analytical methods. 
\begin{conjecture}
\label{conj:three_inst}
The asymptotic expansion of the coefficients in the trans-series~(\ref{mtran}) associated to the third-order problem in~(\ref{ode}) is
\begin{gather} a^{(m)}_{i,j}(n)\sim-(s+1)S_1\sum_{k\ge0}a^{(m+1)}_{i,j}(k)\Gamma(n+\tfrac{35}{12}-k)\nonumber\\
{}+S_1\sum_{k\ge0}\left(4(i+1)a^{(m+1)}_{i+1,j}(k)+6(j+1)a^{(m+1)}_{i,j+1}(k)\right) \Gamma(n-\tfrac{25}{12}-k)\left(\tfrac{21265}{4608}\psi(n-\tfrac{25}{12}-k)+d_1\right)\nonumber\\
{}+\tfrac14S_3\sum_{k\ge0}\left(4(s+1)a^{(m-1)}_{i-1,j}(k)+6(j+1)a^{(m-1)}_{i-2,j+1}(k)\right) (-1)^{n-k}\Gamma(n+\tfrac{25}{12}-k)\nonumber\\
{}-2(s-2i-1)S_3\sum_{k\ge0}a^{(m-1)}_{i,j}(k)(-1)^{n-k}\Gamma(n-\tfrac{35}{12}-k) \left(\tfrac{21265}{4608}\psi(n-\tfrac{35}{12}-k)+f_1\right)\nonumber\\
{}-S_3\sum_{k\ge0}\left(8(i+1)a^{(m-1)}_{i+1,j}(k)+6(j+1)a^{(m-1)}_{i,j+1}(k)\right) (-1)^{n-k}\Gamma(n-\tfrac{95}{12}-k)Q(n-\tfrac{95}{12}-k)\nonumber\\
{}-(f_1-c_1)S_3\sum_{k\ge0}\left(2(i+1)a^{(m-1)}_{i+1,j-1}(k)+6(i+j)a^{(m-1)}_{i,j}(k)\right)(-1)^{n-k}\Gamma(n-\tfrac{35}{12}-k) \label{conj2}\end{gather}
for all integers $m,i,j \geq 0$ 
with $s=m-2i-3j\ge0$, on the understanding that
$a^{(m')}_{i',j'}(n)$ vanishes if $m'-2i'-3j'<0$ or any of $m',i',j'$ is negative,
$\psi(z)$ is the derivative of $\log\Gamma(z)$ and
\begin{equation}Q(z)=\left(\tfrac{21265}{4608}\right)^2\left(\psi^2(z)+\psi^\prime(z)\right)
+2c_1\left(\tfrac{21265}{4608}\right)\psi(z)+c_2\label{Qz}.
\end{equation} 
\end{conjecture}
Note that each term in $Q(z)$ corresponds to a derivative of $\Gamma$ as 
$\psi^2(z) + \psi^\prime(z) = \Gamma'' (z) /\Gamma(z)$.

There are six indeterminate \emph{Stokes constants} in~(\ref{conj2}). These constants are \emph{invariants} of the ODE~(\ref{ode}) and they encode the solution of the \emph{connection problem} to get from a low-order perturbative expansion to a large-order asymptotic  expansion as discussed in Section~\ref{sec:borel}.

We evaluated 1000 digits of each constant. To 50 digits, their values are given by 
\begin{eqnarray} S_1&=&0.087595552909179124483795447421262990627388017406822\ldots\label{S1}\\
d_1&=&-43.332634728250755924500717390319380703460728022278\ldots\label{d1}\\
S_3&=&2.1717853140590990211608601227903892302479464193027\ldots\label{S3}\\
f_1&=&-40.903692509228515003814479126901354785263669553014\ldots\label{f1}\\
c_1&=&-41.031956764302710583921068101545509453704897898188\ldots\label{c1}\\
\frac{c_2}{c_1^2}&=&1.0002016472131992595822805380838324188011572304276\ldots\label{cr} \end{eqnarray}
The Stokes constant $S_1$ in the first line of~(\ref{conj2}) governs
forwards resurgence of non-logarithmic terms. In the second line, $d_1S_1\approx-3.79575$
appears under the logarithmic contribution from $\psi(z)=\log(z)+\bigO\left(\frac{1}{z}\right)$.
Backwards resurgence, with alternating signs, is governed by $S_3$ in the third line.
In the fourth line, $f_1S_3$ appears, under $\psi(z)$. The fifth line involves $c_1$ and $c_2$,
via the abbreviation $Q(z)$, defined in~(\ref{Qz}). The closeness of $c_2$ to $c_1^2$, noted in~(\ref{cr}),
gives a rather good quadratic approximation $Q(z)\approx(\tfrac{21265}{4608}\log(z)+c_1)^2$ at large $z$.
In the final line of~(\ref{conj2}), the small combination $(f_1-c_1)S_3\approx0.278562$
governs the backwards resurgence of the second instanton in the asymptotic expansion
for the coefficients of the third. The value of $S_3$ matches an estimate of this constant 
by Gerald Dunne~\cite{youtubeGerald}, who obtained the first five of its digits by making use of a uniformization map 
combined with Borel--Pad\'e-resummation methods~\cite{Caliceti:2007ra,Costin:2020pcj,Costin:2020hwg}.

The evidence for Conjecture~\ref{conj:three_inst} will be discussed in detail in Section~\ref{sec:resonant}. Before that, we will introduce our methods by applying them to a nonlinear ODE that is 
slightly simpler than~(\ref{ode}). This simplified ODE will feature a resonance between the two solutions of its linearized ODE, but no logarithmic terms. This establishes that the logarithmic terms in~(\ref{ode}) and the similarity to structures that appear in WKB analyses 
of quantum mechanical models cannot be explained merely by the resonant ratio of exponential prefactors in solutions to linearized ODEs. 

\section{Log-free resurgence of a second-order ODE}
\label{sec:logfree}
\subsection{A second-order nonlinear ODE and its trans-series solution}

The second-order differential equation
\begin{equation}
(\td{g}(x)P-1)(\td{g}(x)P-2)\td{g}(x)=1,\quad \text{ where } \quad P=x\left(2x\frac{\dd}{{\dd}x}+1\right),\label{tode}
\end{equation}
with cubic nonlinearity is arguably a natural simplification of~(\ref{ode}). 
The ODE~(\ref{tode}) is the intermediate case between~(\ref{ode}) and the even simpler first-order ODE
$(\td{\td{g}}(x)P-1)\td{\td{g}}(x)=-\frac12$, where $\td{\td{g}}(x)=C(x/2)/x$ and $C(x)$ was the subject of~\cite{BD}. 
Even though the ODE~(\ref{tode}) is simpler than~(\ref{ode}), it is obviously of similar structure. 
The trans-series solution and the all-order asymptotic resurgence analysis of~(\ref{tode}) will be the subject of this section.

As before, we start by solving the linearized homogeneous differential equation for non-per\-tur\-ba\-tive solutions.
We obtain two such solutions:
\begin{equation}
\td{g}(x)=\td{g}_0(x)+\td{\sigma}_k\left(x^{-\frac{9}{4}}\,e^{-\frac{1}{x}}\right)^k \td{h}_k(x)+\bigO(\sigma_k^2),\quad k\in\{1,2\},
\end{equation}
where $\td{g}_0(x)$ is the unique perturbative Frobenius-type solution of~(\ref{tode})
\begin{align} \td{g}_0(x) = \tfrac{1}{2} + \tfrac{3}{8} x + \tfrac{15}{16} x^2 + \tfrac{483}{128} x^3 + \ldots \end{align}
Again, the two non-perturbative solutions $\td{h}_1(x)$ and $\td{h}_2(x)$ to the linearized version of~(\ref{tode}) are \emph{resonant}, as the action of $\td{h}_2(x)$ is twice the action of $\td{h}_1(x)$.
In spite of this resonance, we do not observe logarithmic terms in the trans-series solution of~(\ref{tode}) as we did in~(\ref{tran}--\ref{nsig}). This $\log$-free trans-series  solution of~(\ref{tode}) reads
\begin{gather} \td{g}(x)=\sum_{m=0}^\infty\td{g}_m(x)\td{y}^m,\quad \td{y}=x^{-\frac{9}{4}}\,e^{-\frac{1}{x}},\label{ttran}\\
\td{g}_m(x)=\sum_{j=0}^{\lfloor m/2\rfloor}\sigma_1^{m-2j}\,(\sigma_2x^3)^j\td{T}_{m,j}(x),\quad \td{T}_{m,j}(x)=\sum_{n=0}^\infty\td{a}^{(m)}_{j}(n)x^n.\label{texp} \end{gather}
As was the case in~(\ref{tran}--\ref{nsig}), the second-order ODE~(\ref{tode}) does only fix the values of the coefficients $\td{a}^{(m)}_{j}(n)$ 
up to a few ambiguities. 
Here, we find three such ambiguities.
We resolve two of them by
\begin{gather} \label{tfix1} \td{a}^{(1)}_{0}(0)=-1, \quad \td{a}^{(2)}_{1}(0)=-1, \end{gather}
which normalize $\sigma_1$ and $\sigma_2$. The
third ambiguity comes from the freedom to shift $\sigma_2$ by an arbitrary multiple of $\sigma_1^2$.
We resolve this by
\begin{gather} \label{tfix2} \td{a}^{(2)}_{0}(3)=\tfrac{9855}{512}, \end{gather}
to simplify the asymptotic expansion of the coefficients $\td{a}^{(m)}_{j}(n)$ for large $n$ as much as possible. We will discuss this choice in detail in the next section.

Table~\ref{ttab} gives expansions up to $x^5$ of the 16 terms in~(\ref{ttran}--\ref{texp}) up to $\td{y}^6$ using this normalization.

\begin{table}
{
\small
\begin{gather*}\begin{array}{r|rrrrrr}n&0&1&2&3&4&5\\[2pt]\hline\\[-5pt] \td{a}^{(0)}_{0}(n)&\frac{1}{2}&\frac{3}{8}&\frac{15}{16}&\frac{483}{128}&\frac{5157}{256}&\frac{134343}{1024}\\[2pt] \td{a}^{(1)}_{0}(n)&-1&\frac{33}{16}&\frac{1047}{512}&\frac{70227}{8192}&\frac{25530309}{524288}&\frac{14153452731}{41943040}\\[2pt] \td{a}^{(2)}_{0}(n)&-2&\frac{33}{4}&\frac{3}{64}&\green{\frac{\mathbf{{9855}}}{\mathbf{{512}}}}&\frac{2019789}{16384}&\frac{618176691}{655360}\\[2pt] \td{a}^{(2)}_{1}(n)&-1&\frac{33}{8}&-\frac{525}{128}&\frac{20367}{1024}&\frac{49581}{32768}&\frac{680976531}{1310720}\\[2pt] \td{a}^{(3)}_{0}(n)&-6&\frac{309}{8}&-\frac{11595}{256}&\frac{268575}{4096}&\frac{86978799}{262144}&\frac{59305753719}{20971520}\\[2pt] \td{a}^{(3)}_{1}(n)&-5&\frac{459}{16}&-\frac{20589}{512}&\frac{1020177}{8192}&-\frac{1479015}{524288}&\frac{26407877013}{8388608}\\[2pt] \td{a}^{(4)}_{0}(n)&-\frac{64}{3}&188&-\frac{891}{2}&\frac{1645}{3}&\frac{157613}{192}&\frac{209114347}{23040}\\[2pt] \td{a}^{(4)}_{1}(n)&-24&188&-\frac{5195}{12}&\frac{34183}{36}&-\frac{1953269}{2304}&\frac{1190136713}{69120}\\[2pt] \td{a}^{(4)}_{2}(n)&-2&\frac{35}{2}&-\frac{2731}{48}&\frac{95519}{576}&-\frac{426683}{1024}&\frac{486368029}{184320}\\[2pt] \td{a}^{(5)}_{0}(n)&-\frac{250}{3}&\frac{7475}{8}&-\frac{854895}{256}&\frac{69630643}{12288}&-\frac{675041215}{786432}&\frac{1211128628005}{37748736}\\[2pt] \td{a}^{(5)}_{1}(n)&-\frac{350}{3}&\frac{9425}{8}&-\frac{3034175}{768}&\frac{321421075}{36864}&-\frac{29917076015}{2359296}&\frac{11156692102565}{113246208}\\[2pt] \td{a}^{(5)}_{2}(n)&-\frac{45}{2}&\frac{7537}{32}&-\frac{2820071}{3072}&\frac{390504835}{147456}&-\frac{63051531175}{9437184}&\frac{15960623177125}{452984832}\\[2pt] \td{a}^{(6)}_{0}(n)&-\frac{1728}{5}&\frac{23544}{5}&-\frac{225189}{10}&\frac{4205583}{80}&-\frac{129518877}{2560}&\frac{15136299041}{102400}\\[2pt] \td{a}^{(6)}_{1}(n)&-576&7164&-\frac{64029}{2}&\frac{2654737}{32}&-\frac{76201359}{512}&\frac{78271944319}{122880}\\[2pt] \td{a}^{(6)}_{2}(n)&-180&\frac{4485}{2}&-\frac{341941}{32}&\frac{127558889}{3840}&-\frac{53061308917}{614400}&\frac{5439389708747}{14745600}\\[2pt] \td{a}^{(6)}_{3}(n)&-6&\frac{323}{4}&-\frac{86561}{192}&\frac{39725581}{23040}&-\frac{7036883803}{1228800}&\frac{3546763632961}{147456000} \end{array}\end{gather*}
}
\caption{
Table of the coefficients in the trans-series solution~(\ref{ttran}--\ref{texp}) of~(\ref{tode}). The peculiar choice for the coefficient $\td{a}^{(2)}_{0}(3)$ (in bold and green) leads to a particularly simple form of Conjecture~\ref{tconj}.
}
\label{ttab}
\end{table}

\subsection{Resolution of ambiguities via asymptotic analysis}
The choices of signs in~(\ref{tfix1}) ensure that the coefficients $\td{a}^{(m)}_{j}(n)$ are eventually positive, at large $n$, for all $m\ge2j\ge0$.
The choice for the remaining ambiguity $\td{a}^{(2)}_{0}(3)=\frac{9855}{512}$ in~(\ref{tfix2}) is harder to come by. It is justified by the especially simple \emph{asymptotic} behaviour of the coefficients $\td{a}^{(m)}_{j}(n)$ obtained with it. 
We will discuss the line of thought that leads to this choice in the remainder of this section. Analogous arguments will lead to the ambiguity fixing choices in~(\ref{fix2}) for the full complexity of the solution in~(\ref{tran}--\ref{nsig}) of ODE~(\ref{ode}), 
which we will discuss in Section~\ref{sec:resonant}.

We start by considering the asymptotic behaviour of the coefficients $\td{a}^{(0)}_{0}(n)$ of the perturbative solution $\td{g}_{0}(x)=\sum_{n\ge0}\td{a}^{(0)}_{0}(n)x^n$:
\begin{gather} \td{a}^{(0)}_{0}(n)=\td{S}_1\Gamma(n+\tfrac94)\left(1-\tfrac{33}{16n}+\bigO\left(\tfrac{1}{n^2}\right)\right)\label{ta00l},\\
\td{S}_1=0.17595473991964250815209678804264889548688592517722\ldots\label{tS1} \end{gather}
By the low- and large-order correspondence in the trans-series which has been discussed in Section~\ref{sec:borel}, the asymptotic expansion 
that begins with~(\ref{ta00l}) may be extended, {\em ad libitum}, using
the first-instanton coefficients, $\td{a}^{(1)}_{0}(k)$, in
\begin{equation}
\td{a}^{(0)}_{0}(n)\sim-\td{S}_1\sum_{k\ge0}\td{a}^{(1)}_{0}(k)\Gamma(n+\tfrac{9}{4}-k).\label{ta00}
\end{equation}

To fix the remaining ambiguity, we have to consider the asymptotic expansion 
of the first-instanton coefficients $\td{a}^{(1)}_{0}(n)$ which is governed itself by the second-instanton contribution.
In the second instanton, the second term of $\td{g}_2=\sigma_1^2\td{T}_{2,0}+\sigma_2\td{T}_{2,1}x^3$,
is suppressed by $x^3$. Thus the term of order $x^3$ in $\td{T}_{2,0}$ is ambiguous,
since we may add to the inhomogeneous solution $\td{T}_{2,0}$ any multiple of the
homogeneous instanton solution $\td{T}_{2,1}x^3$.
We resolved this ambiguity by setting $\td{a}^{(2)}_{0}(3)=\tfrac{9855}{512}$ in~(\ref{tfix2}).
{\em Ex post facto}, this ensures that the asymptotic expansion for the first instanton 
\begin{equation}
\td{a}^{(1)}_{0}(n)\sim-2\td{S}_1\sum_{k\ge0}\td{a}^{(2)}_{0}(k)\Gamma(n+\tfrac{9}{4}-k)\label{ta10}
\end{equation}
does not involve the coefficients $\td{a}^{(2)}_{1}(k)$ of the second instanton.
We achieved this empirically, lacking an {\em a priori} method 
to decouple the instantons, by numerically computing the asymptotic expansion of $\td{a}^{(1)}_{0}(n)$ up to sufficiently high order.
This numerical computation revealed the value $\tfrac{9855}{512}$ at the third order of the asymptotic expansion~(\ref{ta10}).

\subsection{Patterns of resurgence}
We proceed to analyze the asymptotic expansions of higher-order instantons.
The asymptotic expansions for the terms in $\td{g}_2=\sigma_1^2\td{T}_{2,0}+\sigma_2\td{T}_{2,1}x^3$ are 
\begin{align} \td{a}^{(2)}_{0}(n)&\sim-3\td{S}_1\sum_{k\ge0}\td{a}^{(3)}_{0}(k)\Gamma(n+\tfrac{9}{4}-k) +\td{S}_2\sum_{k\ge0}\td{a}^{(1)}_{0}(k)(-1)^{n-k}\Gamma(n-\tfrac{9}{4}-k)\label{tFasy}\\
\td{a}^{(2)}_{1}(n)&\sim-\td{S}_1\sum_{k\ge0}\td{a}^{(3)}_{1}(k)\Gamma(n+\tfrac{9}{4}-k) +16\td{S}_2\sum_{k\ge0}\td{a}^{(1)}_{0}(k)(-1)^{n-k}\Gamma(n+\tfrac{3}{4}-k)\label{tCasy} \end{align}
which look forwards, to $\td{g}_3=\sigma_1^3\td{T}_{3,0}+\sigma_1\sigma_2\td{T}_{3,1}x^3$,
and also backwards, to $\td{g}_1=\sigma_1\td{T}_{1,0}$, with alternating signs and an empirical constant
\begin{equation}
\td{S}_2=0.11097873354795693645043942852479413454973815476146\ldots\label{tS3}
\end{equation}
which also appears in the backwards looking terms of
\begin{align} \td{a}^{(3)}_{0}(n)\sim-4\td{S}_1\sum_{k\ge0}\td{a}^{(4)}_{0}(k)\Gamma(n+\tfrac{9}{4}-k) &+2\td{S}_2\sum_{k\ge0}\td{a}^{(2)}_{0}(k)(-1)^{n-k}\Gamma(n-\tfrac{9}{4}-k)\label{tIasy}\\
\td{a}^{(3)}_{1}(n)\sim-2\td{S}_1\sum_{k\ge0}\td{a}^{(4)}_{1}(k)\Gamma(n+\tfrac{9}{4}-k) &+32\td{S}_2\sum_{k\ge0}\td{a}^{(2)}_{0}(k)(-1)^{n-k}\Gamma(n+\tfrac{3}{4}-k)\nonumber\\
&+4\td{S}_2\sum_{k\ge0}\td{a}^{(2)}_{1}(k)(-1)^{n-k}\Gamma(n-\tfrac{9}{4}-k).\label{tEasy} \end{align}
The asymptotic expansions for the terms in $\td{g}_4=\sigma_1^4\td{T}_{4,0} +\sigma_1^2\sigma_2\td{T}_{4,1}x^3+\sigma_2^2\td{T}_{4,2}x^6$ are
\begin{align} \td{a}^{(4)}_{0}(n)\sim&-5\td{S}_1\sum_{k\ge0}\td{a}^{(5)}_{0}(k)\Gamma(n+\tfrac{9}{4}-k) +3\td{S}_2\sum_{k\ge0}\td{a}^{(3)}_{0}(k)(-1)^{n-k}\Gamma(n-\tfrac{9}{4}-k)\label{tKasy}\\
\td{a}^{(4)}_{1}(n)\sim&-3\td{S}_1\sum_{k\ge0}\td{a}^{(5)}_{1}(k)\Gamma(n+\tfrac{9}{4}-k) +48\td{S}_2\sum_{k\ge0}\td{a}^{(3)}_{0}(k)(-1)^{n-k}\Gamma(n+\tfrac{3}{4}-k)\nonumber\\
&+5\td{S}_2\sum_{k\ge0}\td{a}^{(3)}_{1}(k)(-1)^{n-k}\Gamma(n-\tfrac{9}{4}-k)\label{tJasy}\\
\td{a}^{(4)}_{2}(n)\sim&-\td{S}_1\sum_{k\ge0}\td{a}^{(5)}_{2}(k)\Gamma(n+\tfrac{9}{4}-k) +16\td{S}_2\sum_{k\ge0}\td{a}^{(3)}_{1}(k)(-1)^{n-k}\Gamma(n+\tfrac{3}{4}-k).\label{tHasy} \end{align}
\subsection{All-order resurgent asymptotics}
The previous observations lead us to the following conjecture about the large-order behaviour of \emph{all sets of coefficients} $\td{a}^{(m)}_j(n)$ for $n\rightarrow \infty$. 
\begin{conjecture}
\label{tconj}
The asymptotic expansion of the coefficients in the trans-series~(\ref{ttran}--\ref{texp}) associated to the second-order problem in~(\ref{tode}) is
\begin{align} \td{a}^{(m)}_{j}(n)\sim&-\phantom{16}(m-2j+1)\td{S}_1\sum_{k\ge0}\td{a}^{(m+1)}_{j}(k)\Gamma(n+\tfrac{9}{4}-k)\nonumber\\
&+16(m-2j+1)\td{S}_2\sum_{k\ge0}\td{a}^{(m-1)}_{j-1}(k)(-1)^{n-k}\Gamma(n+\tfrac{3}{4}-k)\nonumber\\
&+\phantom{16}(m+2j-1)\td{S}_2\sum_{k\ge0}\td{a}^{(m-1)}_{j}(k)(-1)^{n-k}\Gamma(n-\tfrac{9}{4}-k) \end{align}
for all $m,j$ with $0 \leq 2j \leq m$ on the understanding that $\td{a}^{(m')}_{j'}(n)$ vanishes if $j'<0$ or $m'<2j'$.
\end{conjecture}

Here, in contrast to Conjecture~\ref{conj:three_inst}, only two Stokes constants are sufficient to completely capture the $n\rightarrow \infty$ asymptotic behaviour of the coefficients $\td{a}^{(m)}_{j}(n)$ for all $0 \leq 2j \leq m$.

\begin{table}
{ \small
\begin{gather*}\begin{array}{r|rrrrrr}n&0&1&2&3&4&5\\[2pt]\hline\\[-5pt] \td{a}^{(7)}_{0}(n)&-\frac{67228}{45}&\frac{1438199}{60}&-\frac{275819677}{1920}&\frac{8052601739}{18432}&-\frac{4052588704717}{5898240}&\frac{1561403370991877}{1415577600}\\[2pt] \td{a}^{(7)}_{1}(n)&-\frac{14406}{5}&\frac{5122019}{120}&-\frac{552501215}{2304}&\frac{416889193721}{552960}&-\frac{56661728164699}{35389440}&\frac{40877109335288891}{8493465600}\\[2pt] \td{a}^{(7)}_{2}(n)&-\frac{3773}{3}&\frac{881167}{48}&-\frac{486266039}{4608}&\frac{409923222353}{1105920}&-\frac{120236223547541}{117964800}&\frac{20825178510081761}{5662310400}\\[2pt] \td{a}^{(7)}_{3}(n)&-\frac{637}{6}&\frac{51709}{32}&-\frac{31373293}{3072}&\frac{30785995699}{737280}&-\frac{6647287998833}{47185920}&\frac{31209121062507581}{56623104000}\\[2pt] \td{a}^{(8)}_{0}(n)&-\frac{2097152}{315}&\frac{12910592}{105}&-\frac{18606080}{21}&\frac{211138880}{63}&-\frac{322369024}{45}&\frac{213005543969}{18900}\\[2pt] \td{a}^{(8)}_{1}(n)&-\frac{131072}{9}&\frac{1253376}{5}&-\frac{76564352}{45}&\frac{869171848}{135}&-\frac{435188884}{27}&\frac{67360030181}{1620}\\[2pt] \td{a}^{(8)}_{2}(n)&-8192&\frac{413440}{3}&-\frac{8433232}{9}&\frac{510257878}{135}&-\frac{15238315549}{1350}&\frac{1191247687559}{32400}\\[2pt] \td{a}^{(8)}_{3}(n)&-\frac{3584}{3}&20512&-\frac{442592}{3}&\frac{29788349}{45}&-\frac{3669564881}{1575}&\frac{11441457437239}{1323000}\\[2pt] \td{a}^{(8)}_{4}(n)&-\frac{64}{3}&388&-3079&\frac{126253}{8}&-\frac{326328481}{5040}&\frac{561282005581}{2116800}\\[2pt] \td{a}^{(9)}_{0}(n)&-\frac{1062882}{35}&\frac{177678441}{280}&-\frac{47871004617}{8960}&\frac{696090460671}{28672}&-\frac{600467129297451}{9175040}&\frac{87904551054260979}{734003200}\\[2pt] \td{a}^{(9)}_{1}(n)&-\frac{2598156}{35}&\frac{203935563}{140}&-\frac{1485924345}{128}&\frac{3723462723681}{71680}&-\frac{99869990671719}{655360}&\frac{141892340468228361}{367001600}\\[2pt] \td{a}^{(9)}_{2}(n)&-\frac{255879}{5}&\frac{15643611}{16}&-\frac{19755496863}{2560}&\frac{1463982171813}{40960}&-\frac{1546653558234897}{13107200}&\frac{15539134448654937}{41943040}\\[2pt] \td{a}^{(9)}_{3}(n)&-10935&\frac{3366279}{16}&-\frac{878060655}{512}&\frac{349906464729}{40960}&-\frac{587633538031079}{18350080}&\frac{17882312779069131439}{154140672000}\\[2pt] \td{a}^{(9)}_{4}(n)&-\frac{4131}{8}&\frac{1320219}{128}&-\frac{368401995}{4096}&\frac{32449093361}{65536}&-\frac{1548339054847999}{734003200}&\frac{3507392355461391041}{411041792000} \end{array}\end{gather*}
}
\caption{
Extension of Table~\ref{ttab} for the coefficients in the trans-series solution~(\ref{ttran}--\ref{texp}) of~(\ref{tode}).
}\label{ttab9}
\end{table}

To test Conjecture~\ref{tconj}, we extended Table~\ref{ttab} to order $\td{y}^9$
and the 30 rows of Tables~\ref{ttab} and \ref{ttab9} to $n>100$. Using exact rational values
of $\td{a}^{(m)}_{j}(n)$, we checked the conjecture at 100 digits of precision for $m\le4$, at 70 digits for $m=5$
and at 30 digits for $m=6,7,8$.

In Section~\ref{sec:alien} we will augment some of the empirical observations that were made in previous sections 
to justify Conjecture~\ref{tconj} by an interpretation in terms of \emph{alien calculus}. Before that we will briefly discuss further empirical results which follow from our analysis of~(\ref{tode}) and which enable a partial \emph{trans-asymptotic resummation}.

\subsection{Trans-asymptotics: all-instanton-order results}
Additionally to Conjecture~\ref{tconj} we empirically deduced for the leading coefficients of the trans-series in~(\ref{ttran}--\ref{texp}) that
\begin{equation}
\td{a}^{(m)}_{j}(0)=-(m+2j)\frac{2^{m-2j-1}m^{m-j-2}}{(m-2j)!j!}
\label{atwiddle0}
\end{equation}
for all $m\ge2j\geq 0$.

Setting $j=0$, we have
$\td{a}^{(m)}_{0}(0)=-(-2)^{m-1}w_m$, for $m>0$,
where $w_m=(-m)^{m-1}/m!$ is the coefficient of $x^m$ in the expansion 
of the Lambert-W function on its principal branch. The Lambert function 
$W=\sum_{m=1}^\infty w_mx^m=x-x^2+\frac32x^3+\bigO(x^4)$ 
solves $We^W=x$ for $|x|<\frac{1}{e}$
and appears prominently in a trans-asymptotic analysis of the four-dimensional Yukawa theory renormalon~\cite{BK2,BD}, which is closely related to the renormalon~(\ref{dse}), and also while evaluating  
topological invariants using renormalization methods~\cite{borinsky2020euler} as well as in other expansions of importance with connection to renormalization~\cite{vanBaalen:2008tc,Klaczynski:2013fca,Panzer:2018tvy,Sberveglieri:2020eko}.

\subsection{Alien calculus analysis}
\label{sec:alien}
Alien calculus is an integral part of the resurgence framework~\cite[Ch.~6]{Mitschi:2016fxp} (see also~\cite{borinsky2018generating,borinsky2018graphs} for a simplified version of this technology that works without intricate Borel transform considerations and~\cite{borinsky2018generating} for a direct application to zero-dimensional quantum field theories). In this section we will use this calculus to explain the shapes of equations~(\ref{ta00}--\ref{tFasy}).

We can define a family of differential operators $\mathcal A_{\omega}$ that is indexed by complex numbers $\omega \in \C \setminus \{0 \}$. These operators, \emph{the alien derivatives}, linearly map a formal power series $f(x) \in \C[[x]]$ to another formal power series $(\mathcal A_\omega f)(x) \in \C[[x]]$ that encodes the local expansion of the Borel transform $\mathcal B[A](z)$ near the point $z = \omega$. 

If a factorially divergent power series $f(x) = \sum_{n = 0}^\infty f_n x^n$ has a nontrivial alien derivative $(\mathcal A_{\omega} f)(x) \neq 0$ and $\omega$ is the position of the closest singularity to the origin in the Borel plane of $f(x)$ as discussed in Section~\ref{sec:borel}, then we can translate the knowledge of the alien derivative into information on the large-order behaviour of the coefficients $f_n$ and vice-versa. 
In fact, alien calculus can be seen as an explicit mathematical realisation of the low- and large-order correspondence that has been discussed repeatedly in this article.
For instance, if we have the following power series expansion for $\mathcal A_{\omega} f$,
\begin{gather} \label{defasyA1} (\mathcal A_{\omega} f)(x) = \sum_{k = 0}^\infty c_k x^{k-\beta}, \intertext{ where $\omega$ is the ordinate of the dominant singularity of $\mathcal B[A](z)$, then the asymptotic behaviour of the coefficients of $f(x)$ is given by} \label{defasyA2} f_n \sim \sum_{k \geq 0} c_k \omega^{-n+k-\beta} \Gamma(n -k + \beta) \text{ for } n \rightarrow \infty. \end{gather}
In other words, the coefficients $c_k$ encode both the series representation of the alien derivative $\mathcal A_{\omega} f$ and the large-order behaviour of its coefficients $f_n$.
The utility of this definition of $\mathcal A_{\omega} f$ comes from the fact that the linear alien derivative operators fulfill product and chain rules,
\begin{gather} (\mathcal A_\omega f \cdot g)(x) = g(x) (\mathcal A_\omega f)(x) + f(x) (\mathcal A_\omega g)(x) \\
(\mathcal A_\omega f \circ g)(x) = f'(g(x)) (\mathcal A_\omega g)(x) + e^{-\omega \left( \frac{1}{g(x)} - \frac{1}{x} \right)} (\mathcal A_\omega f)(g(x)), \end{gather}
where we have to assume $g(x) = x + \bigO(x^2)$ for the chain rule to hold. 

The alien derivative does not commute with the ordinary derivative. The commutator reads,
\begin{gather} [ \mathcal A_\omega, \partial_x ] = \frac{\omega}{x^2} \mathcal A_\omega. \label{commut} \end{gather}
These identities are proven in~\cite[Ch.~6]{Mitschi:2016fxp}. See also~\cite{borinsky2018generating} for proofs and derivations in a simplified context.

We can apply the alien derivative operator $\mathcal A_\omega$ to the ODE~(\ref{tode}) from the left and commute it to the right using the the product and the commutation rule. We obtain a linear ODE for $(\mathcal A_{\omega} \td{g})(x)$ of the form,
\begin{gather} p_0(x,\td{g}, \td{g}', \td{g}'',\omega) (\mathcal A_{\omega} \td{g})(x) + p_1(x,\td{g}, \td{g}', \omega) \partial_x (\mathcal A_{\omega} \td{g})(x) + p_2(x,\td{g}) \partial_x^2 (\mathcal A_{\omega} \td{g})(x) = 0, \label{AprimODE} \end{gather}
where $p_0,p_1,p_2$ are polynomials.
The $\omega$-dependence appears due to the commutation rule~(\ref{commut}).
This ODE turns out to only have a power series solution of the form $(\mathcal A_\omega \td{g})(x) = x^{-\beta} ( 1 + \bigO(x))$ if $\omega \in \{1,2\}$. We can conclude that $\mathcal A_\omega \td{g} = 0$ for all $\omega \not \in \{1,2\}$. The solutions for $(\mathcal A_{\omega} \td{g})(x)$ are the formal power series 
\begin{align} (\mathcal A_{1} \td{g})(x) = \td{\mu}_1 &x^{-\frac94} \td{T}_{1,0}(x) \text{ and } \\
(\mathcal A_{2} \td{g})(x) = \td{\mu}_2 &x^{+\frac34} \td{T}_{2,1}(x), \end{align}
where $\td \mu_1$ and $\td \mu_2$ are undetermined integration constants. For the asymptotic behaviour of the original expansion coefficients $\td{a}^{(0)}_{0}(n)$ only the solution for $\omega=1$ is relevant, as it is closest to the origin. The differential equation only fixes the alien derivative $(\mathcal A_1 \td{g})(x)$ up to the overall constant $\mu_1$. We determined the value of this constant numerically. From equation~(\ref{ta00}) and the relationship between asymptotics and alien derivatives~(\ref{defasyA1}--\ref{defasyA2}) it is evident that 
\begin{align} (\mathcal A_1 \td{g})(x) = - \td{S}_1 x^{-\frac94} \td{T}_{1,0}(x) = - \td{S}_1 \sum_{n = 0}^\infty \td{a}^{(1)}_{0}(n) x^{n-\frac94} \end{align}
is the correct expression for the first $\mathcal A_{1}$ derivative of $\td{g}$.

We can repeat the application of alien derivatives. Multiple $\mathcal A_{\omega}$ derivatives with different values of $\omega$ do not commute; they form a \emph{free algebra}. 
For example, we can apply the operator $\mathcal A_1$ twice from the left to the ODE~(\ref{tode}). We get an equation of the form
\begin{gather} \begin{gathered} \label{A1A1ode} p_0(x,\td{g}, \td{g}', \td{g}'', 2) (\mathcal A_{1}^2 \td{g})(x) + p_1(x,\td{g}, \td{g}', 2) \partial_x (\mathcal A_{1}^2 \td{g})(x) + p_2(x,\td{g}) \partial_x^2 (\mathcal A_{1}^2 \td{g})(x) \\
= p_I(x, \td{g}, \td{g}', \td{g}'', (\mathcal A_{1} \td{g}), (\mathcal A_{1} \td{g})', (\mathcal A_{1} \td{g})''), \end{gathered} \end{gather}
which, up to an inhomogeneity that is captured by the polynomial $p_I$, is the same ODE for $(\mathcal A_{1}^2 \td{g})(x)$ as~(\ref{AprimODE}) for $(\mathcal A_\omega \td{g})(x)$ in the $\omega=2$ case. The reason is that $[\mathcal A_1^2, \partial_x] = \frac{2}{x^2} \mathcal A_1^2$. Therefore $\mathcal A_1^2$ behaves similarly to the operator $\mathcal A_2$ for which $[\mathcal A_2, \partial_x] = \frac{2}{x^2} \mathcal A_2$. This leads to a one-parameter family of possible explicit expressions for $(\mathcal A_{1}^2 \td{g})(x)$. Again, the differential equation does not provide sufficient information to completely fix $(\mathcal A_{1}^2 \td{g})(x)$. Empirically, we determined the second alien derivative  $(\mathcal A_{1}^2 \td{g})(x)$ in~(\ref{ta10}). The additional piece of information that $\td{a}^{(2)}_{0}(3)=\tfrac{9855}{512}$ leads to
\begin{align} \label{A1A1g} (\mathcal A_{1}^2 \td{g})(x) = \mathcal A_{1} (\mathcal A_{1} \td{g})(x) = - \td{S}_1 \mathcal A_{1} \left( x^{-\frac94} \td{T}_{1,0}(x) \right) = - \td{S}_1 x^{-\frac94} ( \mathcal A_{1} \td{T}_{1,0} )(x) = 2 \td{S}_1^2 x^{-\frac92} \td{T}_{2,0}(x). \end{align}
Note that the operator $\mathcal A_1$ can be interpreted as the \emph{forwards looking} alien derivative operator as in the illustrative argument from Section~\ref{sec:borel}. Analogously we can interpret $\mathcal A_{-1}$ as the \emph{backwards looking} alien derivative.
An instructive example is the application of the operator $\mathcal A_{-1} \mathcal A_1$ to both sides of the ODE~(\ref{tode}). 
We get
\begin{gather} \begin{gathered} p_0(x,\td{g}, \td{g}', \td{g}'', 0) (\mathcal A_{-1} \mathcal A_{1} \td{g})(x) + p_1(x,\td{g}, \td{g}', 0) \partial_x (\mathcal A_{-1} \mathcal A_{1} \td{g})(x) + p_2(x,\td{g}) \partial_x^2 (\mathcal A_{-1} \mathcal A_{1} \td{g})(x) = 0, \end{gathered} \end{gather}
where we have used that $\mathcal A_{-1} \td{g} = 0$ by~(\ref{AprimODE}) and $-1 \not \in \{1,2\}$. This equation, which is of the form~(\ref{AprimODE}) has no power series solution for $\mathcal A_{-1} \mathcal A_{1} \td{g}$ and we can infer that $\mathcal A_{-1} \mathcal A_{1} \td{g} = 0$. This is in accordance with the missing alternating contribution to the asymptotics of the coefficients $\td{a}^{(1)}_{0}(n)$ in~(\ref{ta10}).

Acting with $\mathcal A_{-1}$ on~(\ref{A1A1ode}), we obtain
\begin{gather} \begin{gathered} p_0(x,\td{g}, \td{g}', \td{g}'', 1) (\mathcal A_{-1} \mathcal A_{1}^2 \td{g})(x) + p_1(x,\td{g}, \td{g}', 1) \partial_x (\mathcal A_{-1} \mathcal A_{1}^2 \td{g})(x) + p_2(x,\td{g}) \partial_x^2 (\mathcal A_{-1} \mathcal A_{1}^2 \td{g})(x) = 0, \end{gathered} \end{gather}
where we have used that $\mathcal A_{-1} \td{g} = \mathcal A_{-1} \mathcal A_{1} \td{g} = 0$ together with the commutator rule for ordinary and alien derivatives.
This homogeneous equation is exactly the ODE in~(\ref{AprimODE}) with $\omega=1$. Therefore, 
\begin{gather} \label{Am1A1A1sol} (\mathcal A_{-1} \mathcal A_{1}^2 \td{g})(x) = \td{\mu}_{-1,1} x^{-\frac94} \td{T}_{1,0}(x), \end{gather}
where the overall constant for the homogeneous solution stays undetermined. 
Translating this into an alien derivative of $\td{T}_{2,0}(x)$ by using~(\ref{A1A1g}) results in 
\begin{gather} \label{Am1A1A1solT} (\mathcal A_{-1} \td{T}_{2,0})(x) = \frac{1}{2S_1^2} \td{\mu}_{-1,1} x^{+\frac94} \td{T}_{1,0}(x), \end{gather}
where we established that indeed the $\td{T}_{1,0}(x)$ sequence reappears in the alternating part of the $\td{T}_{2,0}(x)$ asymptotic expansion in~(\ref{tFasy}).
The exponent of the prefactor $x^{\frac94}$ explains the negative shift of $\frac94$ in the $\Gamma$ function of the alternating part of $(\ref{tFasy})$ due to the alien derivative and asymptotics relation~(\ref{defasyA1}--\ref{defasyA2}). 

We can fix the undetermined number $\td{\mu}_{-1,1}$ numerically by comparing it to the large-order computation in~(\ref{tFasy}) which implies that
\begin{gather} (\mathcal A_{-1} \td{T}_{2,0})(x) = \td{S}_2 (-x)^{+\frac94} \td{T}_{1,0}(x) \end{gather}
where we accounted for the sign $(-1)^{n-k}$ in~(\ref{tFasy}) using the definition~(\ref{defasyA1}--\ref{defasyA2}) of $\mathcal A_\omega$.
Therefore, we have $\td{\mu}_{-1,1} = 2 \td{S}_1^2 \td{S}_2 (-1)^{\frac{9}{4}}$, where the branch of the fourth root in this expression stays undetermined from our simple considerations.

It seems plausible that a complete proof of Conjecture~\ref{tconj} as well as for Conjecture~\ref{conj:three_inst} is achievable using alien calculus methods, but such a proof lies beyond the scope of this article. With this remark we finish our discussion of the alien calculus viewpoint.

\section{Resonant resurgence from a Dyson--Schwinger equation}
\label{sec:resonant}

After this digression on the $\log$-free ODE~(\ref{tode}), we will return to the analysis of~(\ref{ode}), which is associated to the six-dimensional $\phi^3$ theory renormalon singularity described by the Dyson--Schwinger equation~(\ref{dse}) and which features 
the full complexity that is expected from a QFT renormalon. 

In this section we will present the striking evidence for the validity of 
Conjecture~\ref{conj:three_inst} and our reasoning for the peculiar ambiguity fixing choices in~(\ref{fix2}), which involves the large fraction $r_1=\tfrac{32642693907919}{36691771392}$. The analysis and arguments follow similar lines as our analysis of the ODE~(\ref{tode}) in 
the previous section.  

\subsection{Resolution of ambiguities via asymptotic analysis}

To illustrate our process, we introduce a new convention for the $m$-th instanton order expansions in the trans-series solution~(\ref{tran}) for the third-order problem~(\ref{ode}).
In~(\ref{tran}), $g_m$ has contributions proportional to $\sigma_1^{s_1}\sigma_2^{s_2}\sigma_3^{s_3}$
with $m=s_1+2s_2+3s_3$. For $m\le5$ there are 16 such monomials, which we label as follows:
\begin{gather} \begin{array}{rcccccccccccccccc} &A&B&C&D&E&F&G&H&I&J&K&U&V&W&X&Y\\
s_1= &0&1&0&0&1&2&1&0&3&2&4&0&1&2&3&5\\
s_2= &0&0&1&0&1&0&0&2&0&1&0&1&2&0&1&0\\
s_3= &0&0&0&1&0&0&1&0&0&0&0&1&0&1&0&0\\\
s_1+2s_2+3s_3= &0&1&2&3&3&2&4&4&3&4&4&5&5&5&5&5\end{array}\label{dict} \end{gather}
The lexicographic ordering in dictionary~(\ref{dict}) corresponds to the order in which we
investigated progressively more demanding expansions. To each letter, we associate
a formal power series in $x$. 

The perturbative solution $A(x)$ is given by~(\ref{g0}). For the linearized 
non-perturbative solutions in~(\ref{hk}), we use the notation
$h_1(x)=B(x)$, $h_2(x)=x^5C(x)$ and $h_3(x)=x^5D(x)$, with
\begin{align} B(x)&= -1 + \tfrac{97}{48}x + \tfrac{53917}{13824}x^2+\tfrac{3026443}{221184}x^3 + \tfrac{32035763261}{382205952}x^4 +\tfrac{11517422581511}{18345885696}x^5+ \bigO(x^6) \label{Bterms}\\
C(x)&= -1 + \tfrac{151}{24}x - \tfrac{63727}{3456}x^2 + \tfrac{7112963}{82944}x^3 - \tfrac{7975908763x}{23887872}x^4 +\tfrac{517065181955}{191102976}x^5 + \bigO(x^6) \label{Cterms}\\
D(x)&= -1 + \tfrac{227}{48}x + \tfrac{1399}{4608}x^2 + \tfrac{814211}{73728}x^3 + \tfrac{3444654437}{42467328}x^4 +\tfrac{469339409983}{679477248}x^5+ \bigO(x^6). \label{Dterms} \end{align}

Thus $\{A(x),B(x),C(x),D(x)\}$ in~(\ref{g0},\ref{Bterms}--\ref{Dterms})
refer to the perturbative term $g_0=A$ and the instanton terms $\sigma_1yB$, $\sigma_2y^2x^5C$, $\sigma_3y^3x^5D$.
For each letter in the dictionary we follow the convention of~(\ref{g0}), where $A_n$ is the coefficient of $x^n$ in $A(x)$.

In the trans-series we encounter powers of
\begin{equation}
L=\tfrac{21265}{2304}x^5\log(x)\label{Lx}
\end{equation}
with the largest power of $L$ in $g_m$ no greater than $m/2$. Specifically,
\begin{gather} g_0=A,\quad g_1=\sigma_1B,\quad g_2=\sigma_2x^5C+\sigma_1^2(F+CL),\label{g012}\\
g_3=\sigma_3x^5D+\sigma_1\sigma_2x^5E+\sigma_1^3(I+(D+E)L),\label{g3}\\
g_4=\sigma_1\sigma_3x^5G+\sigma_2^2x^{10}H+\sigma_1^2\sigma_2x^5(J+2HL) +\sigma_1^4(K+(G+J)L+HL^2),\label{g4}\\
g_5=\sigma_2\sigma_3x^{10}U+\sigma_1\sigma_2^2x^{10}V+\sigma_1^2\sigma_3x^5(W+UL) +\sigma_1^3\sigma_2x^5(X+(U+2V)L)\nonumber\\
+\sigma_1^5(Y+(W+X)L+(U+V)L^2).\label{g5} \end{gather}
We already resolved one of the three ambiguities by the normalization $B(0)=-1$. 
Two ambiguities remain: we may add to $g_2$ an arbitrary multiple of $\sigma_1^2x^5C$ and we may add
to $g_3$ an arbitrary multiple of $\sigma_1^3x^5D$.

To resolve these ambiguities we again analyze the large-order behaviour of the coefficients of the power series that appear in the trans-series solution.
As in Section~\ref{sec:logfree}, the six terms in the six lines of~(\ref{conj2}) of Conjecture~\ref{conj:three_inst} revealed themselves successively after the analysis of more and more asymptotic expansions of the trans-series coefficients.

We start with the asymptotic expansion of the perturbative coefficients $A_n$ of $g_0$ developed in~\cite{BK2}. 
One readily finds that~\cite{BDM}
\begin{equation}
A_n=S_1\Gamma(n+\tfrac{35}{12})\left(1-\tfrac{97}{48n}+\bigO\left( \tfrac{1}{n^2}\right)\right),
\label{g0lead}
\end{equation}
which is the $m=0$ case of a general \emph{trans-asymptotic} result
\begin{equation}
a^{(m)}_{0,0}(n)=\frac{(2m+2)^m}{m!}S_1\Gamma(n+\tfrac{35}{12})
\left(1-\frac{120m^2+175m+97}{48(m+1)n}+\bigO\left( \frac{1}{n^2}\right)\right),\label{gmlead}
\end{equation}
which we conjecture to hold for all fixed $m$ and which we have checked up to $m=8$.
As in~(\ref{atwiddle0}) we recover the coefficients of the Lambert-W function with a trivial shift. It would be interesting to study the summation of these coefficients over all $m$, but such a study lies beyond the scope of the present article.

The coefficients of $B(x)=\sum_{k\ge0}B_kx^k$, determine the asymptotic expansion
\begin{gather} A_n\sim- S_1\sum_{k\ge0}\Gamma(n+\tfrac{35}{12}-k)B_k,\label{Aasy} \end{gather}
enabling us to determine 3000 digits of $S_1$, by developing 10000 terms of~(\ref{g0})
and 5000 terms of~(\ref{Bterms}).

For the asymptotic expansion of the second-instanton coefficients, we found
\begin{equation} 
C_n\sim-S_1\sum_{k\ge0}E_k\Gamma(n+\tfrac{35}{12}-k)+S_3\sum_{k\ge0}B_k(-1)^{n-k}\Gamma(n+\tfrac{25}{12}-k)\label{Casy}.
\end{equation}
The term with alternating signs in the second sum, which looks backwards to the first instanton, is
suppressed by a factor of $\frac{1}{n^{5/6}}$ relative to the first sum and is multiplied by
the empirically determined constant $S_3$.
The first sum looks forwards, to terms of order $y^3$ in the trans-series, where
\begin{equation}
E(x)=-4 + \tfrac{371}{12}x - \tfrac{111785}{1152}x^2 + \tfrac{8206067}{18432}x^3 - \tfrac{18251431003}{10616832}x^4
+\tfrac{2354671056847}{169869312}x^5+\bigO(x^6)\label{Eterms}
\end{equation}
occurs in the $\sigma_1\sigma_2x^5E$ term of $g_2$.
  It is notable that the third-instanton coefficients $D_k$
  are absent from (\ref{Casy}). This is a consequence of the form of (\ref{tran}--\ref{nsig}).

By developing 5000 terms of $C(x)$, we obtained almost 1500 decimal digits of $S_3$, using about 2500
terms of $E(x)$ and $B(x)$ in optimal truncations of the forwards and backwards looking terms in~(\ref{Casy}).

We continue with the resolution of the two ambiguities to explain our choices in~(\ref{fix2}).
We will have to combine information from multiple asymptotic expansions to obtain a clear picture.

We start with the expansion for the ambiguous log-free term $\sigma_1^2F$ in $g_2$, which we can write with $\alpha$ parameterizing the ambiguity as,
\begin{equation}
F(x)=-2 + \tfrac{49}{6}x + \tfrac{13235}{1728}x^2 +\tfrac{43049}{1728}x^3 + \tfrac{2496477497}{1728}x^4 
+2\alpha x^5+ \bigO(x^6)\label{Fterms}
\end{equation}
bearing in mind that for $n>4$ only the combination $F_n+2\alpha C_{n-5}$ is unambiguous.
Then the asymptotic expansion for the first instanton has the form
\begin{equation}
B_n\sim-2S_1\sum_{k\ge0}F_k\Gamma(n+\tfrac{35}{12}-k)
+4S_1\sum_{k\ge0}C_k\Gamma(n-\tfrac{25}{12}-k)\left(\tfrac{21265}{4608}\psi(n-\tfrac{25}{12}-k)+d_1\right),\label{Basy}
\end{equation}
where $C_k$ multiplies the digamma $\psi$ function
\begin{equation}
\psi(z)=\frac{\Gamma^\prime(z)}{\Gamma(z)}=\log(z)-\frac{1}{2z}+\bigO\left( \frac{1}{z^2} \right).\label{psi} %
\end{equation}
Experiment then assigns a large value to $\alpha+d_1\approx846.31$, which has already been observed in~\cite{BDM}, but offers, as yet, no way
of apportioning that number between $\alpha$ and $d_1$. 

The asymptotic expansion for the third instanton has a backwards constant $S_4$ in
\begin{align} D_n\sim&-S_1\sum_{k\ge0}G_k\Gamma(n+\tfrac{35}{12}-k)-S_4\sum_{k\ge0}C_k(-1)^{n-k}\Gamma(n-\tfrac{35}{12}-k)\label{Dasy}\\
G(x)=&-\tfrac{20}{3}+\tfrac{368}{9}x-\tfrac{9421}{648}x^2+\tfrac{61483}{1296}x^3 +\tfrac{622009525}{1119744}x^4+\tfrac{17261921677}{3359232}x^5 +\bigO(x^6)\label{Gterms}\\
S_4=&~ 0.55712485097773646632802466946834574964057746422381\ldots\label{S4} \intertext{ A second large constant occurs, modulo $\alpha$, in the backwards-looking part of } E_n\sim&-2S_1\sum_{k\ge0}J_k\Gamma(n+\tfrac{35}{12}-k) +8S_1\sum_{k\ge0}H_k\Gamma(n-\tfrac{25}{12}-k)\left(\tfrac{21265}{4608}\psi(n-\tfrac{25}{12}-k)+d_1\right)\nonumber\\
&+2S_3\sum_{k\ge0}F_k(-1)^{n-k}\Gamma(n+\tfrac{25}{12}-k)\nonumber\\
&+4S_3\sum_{k\ge0}C_k(-1)^{n-k}\Gamma(n-\tfrac{35}{12}-k)\left(\tfrac{21265}{4608}\psi(n-\tfrac{35}{12}-k)+e_1\right),\label{Easy}\\
\intertext{where} H(x)=&-2+\tfrac{53}{2}x-\tfrac{72395}{432}x^2+\tfrac{4651117}{5184}x^3 -\tfrac{3511918891}{746496}x^4+\tfrac{259237318621}{8957952}x^5+\bigO(x^6), \label{Hterms} \end{align}
with $\alpha+e_1\approx848.55.$

We expand the ambiguous log-free term $\sigma_1^3I$ in $g_3$, with 
$\beta$ parameterizing the ambiguity, as 
\begin{equation}
I(x)=-6+\tfrac{309}{8}x-\tfrac{8821}{768}x^2+\tfrac{454379}{36864}x^3
+\tfrac{1344528799}{2359296}x^4+2(4\alpha-\beta) x^5+\bigO(x^6)
\label{Iterms}
\end{equation}
bearing in mind that for $n>4$ only the combination $I_n+2(\alpha E_{n-5}-\beta D_{n-5})$ is unambiguous.
The dependence on $\alpha$ comes from an inhomogeneous term, involving products of $g_1$, $g_2$
and their derivatives.
Then $\beta$ parameterizes an undetermined
homogeneous contribution to $I(x)$. We investigated the asymptotic behaviours
of $F_n$ and $I_n$ with $\alpha=\beta=0$. Our empirical results involved a large rational number
\begin{equation}
r_1=\tfrac{32642693907919}{36691771392}\approx889.646\ldots\label{r1}
\end{equation}
whose several appearances were confirmed at more than 200 decimal digits of precision. We found that the
explicit appearance of $r_1$ was removed by choosing $\alpha=\beta=r_1$. This gives
$a^{(2)}_{0,0}(5)=2r_1$ and $a^{(3)}_{0,0}(5)=6r_1$, in~(\ref{fix2}), and removes $r_1$ from
\begin{align} F_n\sim&-3S_1\sum_{k\ge0}I_k\Gamma(n+\tfrac{35}{12}-k)\nonumber\\
&+2S_1\sum_{k\ge0}(3D_k+2E_k)\Gamma(n-\tfrac{25}{12}-k)\left(\tfrac{21265}{4608}\psi(n-\tfrac{25}{12}-k)+d_1\right)\nonumber\\
&-2S_3\sum_{k\ge0}B_k(-1)^{n-k}\Gamma(n-\tfrac{35}{12}-k)\left(\tfrac{21265}{4608}\psi(n-\tfrac{35}{12}-k)+f_1\right)\label{Fasy} \end{align} 
with an empirical relation
\begin{equation}
3S_4=4(f_1-e_1)S_3.\label{f1rel}
\end{equation}
By this method, we arrived at a resolution of ambiguities that produces comparatively simple  
asymptotic expansions in Conjecture~\ref{conj:three_inst}, which do not explicitly involve $r_1$. 

We proceed to collect further evidence for the validity of Conjecture~\ref{conj:three_inst}.

\subsection{Patterns of resurgence}

Asymptotic expansions of the coefficients in $g(x) = \sum_{m \geq 0} g_m(x) y^m$ at order $y^3$ involves the expansions of
\begin{align} J(x)&=-\tfrac{52}{3}+\tfrac{1567}{9}x-\tfrac{442379}{648}x^2+\tfrac{7216345}{2592}x^3 -\tfrac{12483431383}{1119744}x^4 +\bigO(x^5)\label{Jterms}\\
K(x)&=-\tfrac{64}{3}+\tfrac{1702}{9}x-\tfrac{188909}{648}x^2 -\tfrac{214877}{2592}x^3 +\tfrac{ 1175872537}{559872}x^4 +\bigO(x^5)\label{Kterms} \end{align}
which appear in 
\begin{align} I_n\sim&-4S_1\sum_{k\ge0}K_k\Gamma(n+\tfrac{35}{12}-k)\nonumber\\
&+2S_1\sum_{k\ge0}(3G_k+2J_k)\Gamma(n-\tfrac{25}{12}-k)\left(\tfrac{21265}{4608}\psi(n-\tfrac{25}{12}-k)+d_1\right)\nonumber\\
&-4S_3\sum_{k\ge0}F_k(-1)^{n-k}\Gamma(n-\tfrac{35}{12}-k)\left(\tfrac{21265}{4608}\psi(n-\tfrac{35}{12}-k)+f_1\right)\nonumber\\
&-8S_3\sum_{k\ge0}C_k(-1)^{n-k}\Gamma(n-\tfrac{95}{12}-k)Q(n-\tfrac{95}{12}-k),\label{Iasy}\\
Q(z)=&\left(\tfrac{21265}{4608}\right)^2\left(\psi^2(z)+\psi^\prime(z)\right) +2c_1\left(\tfrac{21265}{4608}\right)\psi(z)+c_2\label{Qagain}, \end{align}
which is also free of $r_1$. Moreover, we found the empirical relation
\begin{equation}S_4=2S_3(c_1-f_1).\label{S4rel}\end{equation}

At order $y^5$, we developed 200 terms of each of the 5 expansions
\begin{align} U(x)&= -\tfrac{13}{3}+\tfrac{7529}{144}x-\tfrac{10635103}{41472}x^2+\tfrac{2075033425}{1990656}x^3 -\tfrac{5055590564695}{1146617856}x^4+\bigO(x^5)\label{Uterms}\\
V(x)&= -\tfrac{109}{6}+\tfrac{77129}{288}x-\tfrac{150627967}{82944}x^2+\tfrac{39350259505}{3981312}x^3 -\tfrac{119208170587255}{2293235712}x^4+\bigO(x^5)\label{Vterms}\\
W(x)&= -\tfrac{110}{3}+\tfrac{21193}{72}x-\tfrac{7613617}{20736}x^2+\tfrac{3848375}{36864}x^3 +\tfrac{1683959379395}{573308928}x^4+\bigO(x^5)\label{Wterms}\\
X(x)&= -80+\tfrac{18073}{18}x-\tfrac{2117255}{432}x^2+\tfrac{4766702935}{248832}x^3 -\tfrac{1833416477245}{23887872}x^4+\bigO(x^5)\label{Xterms}\\
Y(x)&= -\tfrac{250}{3}+\tfrac{67925}{72}x-\tfrac{54524255}{20736}x^2+\tfrac{265397983}{331776}x^3 +\tfrac{6083187427417}{573308928}x^4+\bigO(x^5)\label{Yterms} \end{align} 
in order to have good control of the forwards looking non-alternating parts of the asymptotic expansions
of $\{G_n,H_n,J_n,K_n\}$, from order $y^4$. We found that
\begin{align} G_n\sim&-2S_1\sum_{k\ge0}W_k\Gamma(n+\tfrac{35}{12}-k)\nonumber\\
&+4S_1\sum_{k\ge0}U_k\Gamma(n-\tfrac{25}{12}-k)\left(\tfrac{21265}{4608}\psi(n-\tfrac{25}{12}-k)+d_1\right)\nonumber\\
&-S_4\sum_{k\ge0}(3D_k+E_k)(-1)^{n-k}\Gamma(n-\tfrac{35}{12}-k)\label{Gasy} \\
H_n\sim&-S_1\sum_{k\ge0}V_k\Gamma(n+\tfrac{35}{12}-k)\nonumber\\
&+\tfrac12S_3\sum_{k\ge0}(3D_k+2E_k)(-1)^{n-k}\Gamma(n+\tfrac{25}{12}-k)\label{Hasy}  \end{align}
\begin{align} J_n\sim&-3S_1\sum_{k\ge0}X_k\Gamma(n+\tfrac{35}{12}-k)\nonumber\\
&+2S_1\sum_{k\ge0}(3U_k+4W_k)\Gamma(n-\tfrac{25}{12}-k)\left(\tfrac{21265}{4608}\psi(n-\tfrac{25}{12}-k)+d_1\right)\nonumber\\  &+3S_3\sum_{k\ge0}I_k(-1)^{n-k}\Gamma(n+\tfrac{25}{12}-k)\nonumber\\
&+2S_3\sum_{k\ge0}E_k(-1)^{n-k}\Gamma(n-\tfrac{35}{12}-k)\left(\tfrac{21265}{4608}\psi(n-\tfrac{35}{12}-k)+2e_1-f_1\right) \label{Jasy} \\
K_n\sim&-5S_1\sum_{k\ge0}Y_k\Gamma(n+\tfrac{35}{12}-k) \nonumber\\
&+2S_1\sum_{k\ge0}(3W_k+2X_k)\Gamma(n-\tfrac{25}{12}-k)\left(\tfrac{21265}{4608}\psi(n-\tfrac{25}{12}-k)+d_1\right) \nonumber\\  &-6S_3\sum_{k\ge0}I_k(-1)^{n-k}\Gamma(n-\tfrac{35}{12}-k)\left(\tfrac{21265}{4608}\psi(n-\tfrac{35}{12}-k)+f_1\right) \nonumber\\
&-2S_3\sum_{k\ge0}(3D_k+4E_k)(-1)^{n-k}\Gamma(n-\tfrac{95}{12}-k)Q(n-\tfrac{95}{12}-k) \label{Kasy} \end{align}
with $\{c_1,c_2\}$ appearing in
the final line, which uses the abbreviation~(\ref{Qz}).

In conclusion, no new constant emerges from the asymptotic expansion of coefficients at order $y^4$, 
neither looking forwards, to order $y^5$, nor backwards, to order $y^3$.

With these findings we may return to the convention for the coefficients in~(\ref{mtran}) and 
conclude that  the number of  monomials in~(\ref{tran}) at order $y^m$
is the integer nearest to $(m+3)^2/12$, which is 7 in the case of $m=6$, with 7 formal power series in
\begin{gather} g_6=\widehat{\sigma}_2^3T^{(6)}_{3,0}x^{15} +\left(\sigma_1^2\widehat{\sigma}_2^2T^{(6)}_{2,0} +\sigma_1\widehat{\sigma}_2\widehat{\sigma}_3T^{(6)}_{1,1} +\widehat{\sigma}_3^2T^{(6)}_{0,2}\right)x^{10}\nonumber\\
+\left(\sigma_1^4\widehat{\sigma}_2T^{(6)}_{1,0} +\sigma_1^3\widehat{\sigma}_3T^{(6)}_{0,1}\right)x^5 +\sigma_1^6T^{(6)}_{0,0} \label{g6} \end{gather}
and $T^{(6)}_{i,j}=\sum_{n\ge0}a^{(6)}_{i,j}(n)x^n$ expanded to order $x^5$ in Table~\ref{tab6}.

\begin{table}
\begin{gather*}\begin{array}{r|rrrrrr}n&0&1&2&3&4&5\\[2pt]\hline\\[-5pt] a^{(6)}_{3,0}(n)&-6&\frac{487}{4}&-\frac{229217}{192}&\frac{65775923}{7680}&-\frac{600452444947}{11059200}&\frac{457959210595637}{1327104000}\\[2pt] a^{(6)}_{2,0}(n)&-126&\frac{8463}{4}&-\frac{5127107}{320}&\frac{3486217751}{38400}&-\frac{8869377115513}{18432000}&\frac{600844499892388373}{212336640000}\\[2pt] a^{(6)}_{1,1}(n)&-52&\frac{4229}{6}&-\frac{5460533}{1440}&\frac{2636883739}{172800}&-\frac{581594671123}{9216000}&\frac{3716005285749107}{9953280000}\\[2pt] a^{(6)}_{0,2}(n)&-2&\frac{245}{12}&-\frac{157601}{2880}&\frac{19083283}{345600}&\frac{666546523}{6144000}&\frac{23521502839379}{19906560000}\\[2pt] a^{(6)}_{1,0}(n)&-384&5796&-\frac{825223}{24}&\frac{401400947}{2880}&-\frac{125918348441}{230400}&\frac{146245935305193763}{47775744000}\\[2pt] a^{(6)}_{0,1}(n)&-192&1950&-\frac{53899}{12}&\frac{4801723}{2880}&\frac{977907979}{57600}&\frac{12393093312450199}{95551488000}\\[2pt] a^{(6)}_{0,0}(n)&-\frac{1728}{5}&\frac{23832}{5}&-\frac{384519}{20}&\frac{1599631}{80}&\frac{52207793}{960}&\frac{1670028654245}{10616832} \end{array}\end{gather*}
\caption{
Coefficients $a^{(6)}_{i,j}(n)$ in the trans-series solution~(\ref{mtran}).
}
\label{tab6}
\end{table}
We expanded~(\ref{g6}) to order $x^{150}$, finding the asymptotic expansions
\begin{align} a^{(5)}_{1,1}(n)\sim&-S_1\sum_{k\ge0}a^{(6)}_{1,1}(k)\Gamma(n+\tfrac{35}{12}-k)\nonumber\\
&+S_3\sum_{k\ge0}a^{(4)}_{0,1}(k)(-1)^{n-k}\Gamma(n+\tfrac{25}{12}-k)\nonumber\\
&-2S_4\sum_{k\ge0}a^{(4)}_{2,0}(k)(-1)^{n-k}\Gamma(n-\tfrac{35}{12}-k), \label{Uasy} \end{align}
\begin{align}  a^{(5)}_{2,0}(n)\sim&-2S_1\sum_{k\ge0}a^{(6)}_{2,0}(k)\Gamma(n+\tfrac{35}{12}-k)\nonumber\\
&+12S_1\sum_{k\ge0}a^{(6)}_{3,0}(k)\Gamma(n-\tfrac{25}{12}-k)\left(\tfrac{21265}{4608}\psi(n-\tfrac{25}{12}-k)+d_1\right)\nonumber\\  &+\tfrac12S_3\sum_{k\ge0}\left(4a^{(4)}_{1,0}(k)+3a^{(4)}_{0,1}(k)\right)(-1)^{n-k}\Gamma(n+\tfrac{25}{12}-k)\nonumber\\
&+8S_3\sum_{k\ge0}a^{(4)}_{2,0}(k)(-1)^{n-k}\Gamma(n-\tfrac{35}{12}-k)\left(\tfrac{21265}{4608}\psi(n-\tfrac{35}{12}-k)+e_1\right),\label{Vasy}   \\
a^{(5)}_{0,1}(n)\sim&-3S_1\sum_{k\ge0}a^{(6)}_{0,1}(k)\Gamma(n+\tfrac{35}{12}-k)\nonumber\\
&+4S_1\sum_{k\ge0}\left(a^{(6)}_{1,1}(k)+3a^{(6)}_{0,2}(k)\right)\Gamma(n-\tfrac{25}{12}-k)\left(\tfrac{21265}{4608}\psi(n-\tfrac{25}{12}-k)+d_1\right)\nonumber\\  &-2S_3\sum_{k\ge0}a^{(4)}_{0,1}(k)(-1)^{n-k}\Gamma(n-\tfrac{35}{12}-k)\left(\tfrac{21265}{4608}\psi(n-\tfrac{35}{12}-k)+3f_1-2e_1\right)\nonumber\\
&-S_4\sum_{k\ge0}a^{(4)}_{1,0}(k)(-1)^{n-k}\Gamma(n-\tfrac{35}{12}-k), \label{Wasy}   \\
a^{(5)}_{1,0}(n)\sim&-4S_1\sum_{k\ge0}a^{(6)}_{1,0}(k)\Gamma(n+\tfrac{35}{12}-k)\nonumber\\
&+2S_1\sum_{k\ge0}\left(4a^{(6)}_{2,0}(k)+3a^{(6)}_{1,1}(k)\right)\Gamma(n-\tfrac{25}{12}-k)\left(\tfrac{21265}{4608}\psi(n-\tfrac{25}{12}-k)+d_1\right)\nonumber\\  &+4S_3\sum_{k\ge0}a^{(4)}_{0,0}(k)(-1)^{n-k}\Gamma(n+\tfrac{25}{12}-k)\nonumber\\
&-16S_3\sum_{k\ge0}a^{(4)}_{2,0}(k)(-1)^{n-k}\Gamma(n-\tfrac{95}{12}-k)Q(n-\tfrac{95}{12}-k)\nonumber\\
&-3S_4\sum_{k\ge0}a^{(4)}_{1,0}(k)(-1)^{n-k}\Gamma(n-\tfrac{35}{12}-k), \label{Xasy} \end{align}
\begin{align}  a^{(5)}_{0,0}(n)\sim&-6S_1\sum_{k\ge0}a^{(6)}_{0,0}(k)\Gamma(n+\tfrac{35}{12}-k)\nonumber\\
&+2S_1\sum_{k\ge0}\left(2a^{(6)}_{1,0}(k)+3a^{(6)}_{0,1}(k)\right)\Gamma(n-\tfrac{25}{12}-k)\left(\tfrac{21265}{4608}\psi(n-\tfrac{25}{12}-k)+d_1\right)\nonumber\\  &-8S_3\sum_{k\ge0}a^{(4)}_{0,0}(k)(-1)^{n-k}\Gamma(n-\tfrac{35}{12}-k)\nonumber\\
&-2S_3\sum_{k\ge0}\left(4a^{(4)}_{1,0}(k)+3a^{(4)}_{0,1}(k)\right)(-1)^{n-k}\Gamma(n-\tfrac{95}{12}-k)Q(n-\tfrac{95}{12}-k). \label{Yasy} \end{align}
At this stage, we sought to consolidate our findings, arriving at the tentative Ansatz~(\ref{conj2}),
in which the factors $(s+1)$, $(i+1)$ or $(j+1)$ appear when the values of $s=m-2i-3j$, $i$ or $j$
on the left-hand side are increased by unity in the coefficients on the right-hand side. There remained
some doubt about the factors multiplying $a^{(m-1)}_{i,j}(k)$, which appears in both the fourth
and sixth lines of~(\ref{conj2}). 
To strengthen the case for Conjecture~\ref{conj:three_inst}, we checked it at $m=6,7,8$, as follows.

\begin{table}
{ \small
\begin{gather*}\begin{array}{r|rrrrrr}n&0&1&2&3&4&5\\[2pt]\hline\\[-5pt] a^{(7)}_{3,0}(n)&-\frac{259}{3}&\frac{271373}{144}&-\frac{4054166413}{207360}&\frac{7200569175173}{49766400}&-\frac{133224071564501669}{143327232000}&\frac{67641347825361224921}{11466178560000}\\[2pt] a^{(7)}_{2,1}(n)&-\frac{119}{6}&\frac{36563}{96}&-\frac{1382103449}{414720}&\frac{2053997468369}{99532800}&-\frac{32652433354365457}{286654464000}&\frac{45684245159586373439}{68797071360000}\\[2pt] a^{(7)}_{2,0}(n)&-\frac{7175}{9}&\frac{6575987}{432}&-\frac{80713584797}{622080}&\frac{38740153206949}{49766400}&-\frac{1797176877881965801}{429981696000}&\frac{4952153931106909139999}{206391214080000}\\[2pt] a^{(7)}_{1,1}(n)&-\frac{3878}{9}&\frac{1433203}{216}&-\frac{12604214377}{311040}&\frac{4184421797869}{24883200}&-\frac{146004884277086381}{214990848000}&\frac{388039641121215199669}{103195607040000}\\[2pt] a^{(7)}_{0,2}(n)&-\frac{266}{9}&\frac{74899}{216}&-\frac{356314747}{311040}&\frac{31544236729}{24883200}&\frac{366558672023629}{214990848000}&\frac{851081516906500277}{51597803520000}\\[2pt] a^{(7)}_{1,0}(n)&-\frac{85064}{45}&\frac{9001741}{270}&-\frac{18244366679}{77760}&\frac{257235420119}{248832}&-\frac{43100485872895199}{10749542400}&\frac{105252027010556594701}{5159780352000}\\[2pt] a^{(7)}_{0,1}(n)&-\frac{8918}{9}&\frac{2656633}{216}&-\frac{2625708449}{62208}&\frac{186761126831}{4976640}&\frac{4438066964569991}{42998169600}&\frac{2819066592764568749}{5159780352000}\\[2pt] a^{(7)}_{0,0}(n)&-\frac{67228}{45}&\frac{13126267}{540}&-\frac{3992600605}{31104}&\frac{581323789837}{2488320}&\frac{810130476586739}{4299816960}&\frac{49672894752593599}{412782428160} \end{array}\end{gather*}
}
\caption{
Coefficients $a^{(7)}_{i,j}(n)$ in the trans-series solution~(\ref{mtran}).
}
\label{tab7}
\end{table}
To control resurgence from order $y^6$ to order $y^7$, we extended our coefficient database to include Table~\ref{tab7} 
and obtained all 8 terms in $g_7(x)$ up to order $x^{100}$. 
Then we checked Conjecture~\ref{conj:three_inst} in the 7 cases with $m=6$,
at precisions ranging between 38 and 42 decimal digits.

\begin{table}
{ \small
\begin{gather*}\begin{array}{r|rrrrrr}n&0&1&2&3&4&5\\[2pt]\hline\\[-5pt] a^{(8)}_{4,0}(n)&-\frac{64}{3}&\frac{5252}{9}&-\frac{629393}{81}&\frac{140050183}{1944}&-\frac{169159145641}{306180}&\frac{223796627751307}{57153600}\\[2pt] a^{(8)}_{1,2}(n)&-\frac{176}{9}&\frac{9292}{27}&-\frac{6074089}{2430}&\frac{2995201043}{255150}&-\frac{31345393168001}{642978000}&\frac{2785091374314479}{11252115000}\\[2pt] a^{(8)}_{3,0}(n)&-\frac{7472}{9}&\frac{532612}{27}&-\frac{535414159}{2430}&\frac{218106415984}{127575}&-\frac{7226064335619611}{642978000}&\frac{1097978271769426957123}{15362887680000}\\[2pt] a^{(8)}_{2,1}(n)&-\frac{3104}{9}&\frac{193168}{27}&-\frac{81079616}{1215}&\frac{108766001009}{255150}&-\frac{760156349159689}{321489000}&\frac{305892983098665409}{22504230000}\\[2pt] a^{(8)}_{2,0}(n)&-\frac{43520}{9}&\frac{312116}{3}&-\frac{1208665517}{1215}&\frac{6525798233137}{1020600}&-\frac{11276241916699183}{321489000}&\frac{62481517793880963279701}{316036546560000}\\[2pt] a^{(8)}_{1,1}(n)&-3072&\frac{1447448}{27}&-\frac{151645958}{405}&\frac{846230018017}{510300}&-\frac{353924320643951}{53581500}&\frac{5376881501091909491491}{158018273280000}\\[2pt] a^{(8)}_{0,2}(n)&-\frac{2560}{9}&\frac{103828}{27}&-\frac{19274929}{1215}&\frac{7847142763}{340200}&\frac{556843904228}{40186125}&\frac{42888089323075108997}{316036546560000}\\[2pt] a^{(8)}_{1,0}(n)&-\frac{425984}{45}&\frac{8593408}{45}&-\frac{379432856}{243}&\frac{11137109981}{1458}&-\frac{78987170399093}{2624400}&\frac{1242513062280803219}{8817984000}\\[2pt] a^{(8)}_{0,1}(n)&-\frac{229376}{45}&\frac{10145792}{135}&-\frac{419925544}{1215}&\frac{433075139}{810}&\frac{1235117514193}{2624400}&\frac{7637418200035853}{4408992000}\\[2pt] a^{(8)}_{0,0}(n)&-\frac{2097152}{315}&\frac{117997568}{945}&-\frac{6932873728}{8505}&\frac{2594246072}{1215}&-\frac{412759751107}{918540}&-\frac{8349936031871}{3149280} \end{array}\end{gather*}
}
\caption{
Coefficients $a^{(8)}_{i,j}(n)$ in the trans-series solution~(\ref{mtran}).
}
\label{tab8}
\end{table}

To control resurgence from order $y^7$ to order $y^8$,  we extended 
our coefficient database to include Table~\ref{tab8} 
and obtained all 10 terms in $g_8(x)$ up to order $x^{55}$. 
Then we checked Conjecture~\ref{conj:three_inst} in the 8 cases with $m=7$,
at precisions ranging between 22 and 27 decimal digits.

Finally we developed all 12 terms in $g_9(x)$ up to order $x^{30}$ and checked Conjecture~\ref{conj:three_inst} in the 10 cases with $m=8$,
at precisions ranging between 7 and 17 decimal digits.

After this labour, we are well persuaded by the data that the conjecture holds.

Before we conclude, we will provide some details on our computational methodology for the explicit computation of the coefficients $a^{(m)}_{i,j}(n)$ up to sufficiently large values of $n$.

\subsection{Efficient calculation of expansion coefficients}

To investigate asymptotic behaviour of the coefficients $a^{(m)}_{i,j}(n)$ in~(\ref{mtran}), we developed at least 4000 terms of each of the 11 rational
sequences involved up to order $m \leq 4$, 200 terms of 5 sequences at order $m=5$,
150 terms of 7 sequences at order $m=6$, 100 terms of terms of 8 sequences at order $m=7$,
55 terms of 10 sequences at order $m=8$  and 30 terms of 12 sequences at order $m=9$.

To do this, we had to solve linearized inhomogeneous third-order differential equations, with coefficients that are cubic
in $g_0$ and its derivatives. The inhomogeneous terms are quartics in lower terms of the trans-series
and their derivatives. At order $m=8$, they depend quartically on $\log(x)$.
This complexity is compounded by the differing actions of the operator ${\cal D}=x\tfrac{\dd}{\dd x}$
on $x$, $y$ and $\log(x)$. 

In this work, we were greatly aided by the {\texttt{diffop}} operator of Pari-GP~\cite{GP},
which enabled us to prepare problems symbolically, before performing series expansions and
solving recursions. 
When one has a pair of functions in play, the GP process 
\begin{verbatim}
  fvec = [x,y,logx, u0,u1,u2, v0,v1,v2];
  dvec = [x,(1/x-35/12)*y,1, u1,u2,u3, v1,v2,v3];
  Dz = diffop(z,fvec,dvec);
\end{verbatim}
will perform the action of ${\cal D}$, symbolically, on any rational function $z$ of
$\{x,y,\log(x)\}$, $u_0(x)$ and its derivatives, $v_0(x)$ and its derivatives, on the understanding that
$u_{k+1}$ and $v_{k+1}$ represent the actions of ${\cal D}$ on $u_k$ and $v_k$ and that 
${\cal D}$ is applied no more than three times to $u_0$ and $v_0$. 

After obtaining these symbolic equations, we then took care to process products of power series
in a way that minimizes the many multiplications of series resulting from quartic nonlinearity.
To handle logarithmic dependence, we exploited simplifications that arose from having solved previous 
problems with lesser powers of logs, following the example of Frobenius. This preparation was particularly intricate in the case
of the $\sigma_1^4$ term in~(\ref{tran}--\ref{nsig}). Thereafter, we were able to develop 4000 exact terms of the sequence
$a^{(4)}_{0,0}(n)$ in 6 hours on a laptop.

\section{Conclusion}

We performed an exhaustive resurgence analysis of the nonlinear ODE~(\ref{ode}) which describes the renormalon contribution~(\ref{dse}) to the $\phi^3$ theory field anomalous dimension in six dimensions. The coefficients $A_n = a^{(0)}_{0,0}(n)$ of the perturbative solution of this differential equation have a large-order behaviour which is encoded by the first-instanton contribution to the trans-series solution~(\ref{tran}--\ref{nsig}) with coefficients $a^{(1)}_{0,0}(n)$. The large-order behaviour of these coefficients $a^{(1)}_{0,0}(n)$ is encoded by a combination of the sequences $a^{(2)}_{0,0}(n)$ and $a^{(2)}_{1,0}(n)$, the next terms in the trans-series~(\ref{mtran}). 
This low- and large-order relationship persists 
for all higher-order terms in the trans-series solution to~(\ref{ode}). 
Our analysis was greatly facilitated by the discovery 
of the compact form of the trans-series~(\ref{tran}--\ref{nsig}), which tamed the dependence on the logarithmic terms, and by ambiguity fixing choices, which the asymptotic expansions and were determined by our asymptotic analysis. The compact form of the trans-series solution~(\ref{tran}--\ref{nsig}) for~(\ref{ode}) was the key to go beyond the previous work on this renormalon~\cite{BDM} and constitutes our first main result.

The role of the ambiguities that appear due to resonances of the various expansions turned out to be quite intricate. By comparing multiple large-order expansions, we were able to  empirically find the particularly favourable ambiguity resolving constant $r_1=\frac{32642693907919}{36691771392}$. This numerically determined value reduced the size of our expressions and the magnitude of the presumably transcendental constants significantly. 
We remark that the existence of the compact representation~(\ref{tran}--\ref{nsig}), which hints for a cancellation mechanism phenomenon that ensures reality after resummation, and the role of the rather peculiar constant $r_1$ deserve an explanation that lies beyond the scope of this article.

Our second main result is Conjecture~\ref{conj:three_inst}: the well-tested formula~(\ref{conj2}) for the large-$n$ asymptotics of any sequence $a^{(m)}_{i,j}(n)$ that appears in the trans-series solution of~(\ref{ode}). This large-$n$ behaviour is entirely determined by a linear combination of \emph{nearby} sequences $a^{(m')}_{i',j'}$ which also contribute to the trans-series expansion. This linear combination involves six \emph{Stokes constants} which appear to be transcendental and which we determined numerically. We therefore solved the \emph{connection problem} of matching the low-order with the large-order expansion for the ODE~(\ref{ode}).

To illustrate our process and to showcase the special role of the logarithmic terms in the initial problem, 
we also analyzed 
the related but simpler ODE~(\ref{tode}). This simpler ODE's trans-series solution~(\ref{ttran}--\ref{texp}) lacks the intricate logarithmic terms 
that feature in 
the original ODE's trans-series solution~(\ref{tran}--\ref{nsig}), but inherits the property of a resonance in the linearized non-perturbative solution space. The observation that the simpler ODE~(\ref{tode}) does not feature logarithmic terms in spite of such resonances proves that the origin of the logarithmic contributions is quite peculiar and that such resonances do not necessarily imply the existence of logarithmic terms in the trans-series solution. It would be highly beneficial to refine the notion of \emph{resonant instanton structure} and characterize ODEs in whose trans-series solution logarithmic terms appear. 

With Conjecture~\ref{tconj} we gave the simpler counterpart of Conjecture~\ref{conj:three_inst}: a well-tested formula for the large-$n$ behaviour of the coefficients of the trans-series solution of the simpler ODE~(\ref{tode}).

In Section~\ref{sec:alien} we illuminated the case for Conjecture~\ref{tconj} using alien calculus. It might be possible to prove Conjectures~\ref{conj:three_inst}--\ref{tconj} and to explain the compact trans-series solutions~(\ref{tran}--\ref{nsig}) and~(\ref{ttran}--\ref{texp}) by a more sophisticated application of similar alien calculus reasoning. We leave this for a future work.

The analysis of a similar renormalon in Yukawa theory~\cite{BD} was greatly facilitated 
by the fact that the associated perturbative solution has a sound combinatorial interpretation as the generating function of connected chord diagrams. This interpretation led to an explicit \emph{all-order} trans-series solution of the associated ODE in terms of a certain generating function. The combinatorial interpretation and its relation to the asymptotic expansion was studied further in~\cite{Mahmoud:2020vww,mahmoud2020asymptotic,Mahmoud:2020qbb,AssemMahmoud:2020rvk}. Unfortunately, it seems unlikely that a similar combinatorial interpretation exists for the solution of~(\ref{ode}). Assuming the validity of Conjecture~\ref{conj:three_inst}, we were able to obtain almost the same level of understanding of the $\phi^3$ theory renormalon~(\ref{dse}) as~\cite{BD} achieved for the simpler Yukawa renormalon. We gained complete control over the asymptotic connections between the different trans-series coefficients, even though we do not have an explicit formula for six Stokes constants, which had to be determined numerically.

For both ODEs~(\ref{ode}) and~(\ref{tode}) we found trans-asymptotic results that involve coefficients of the Lambert-$W$ function. 
These might be of help in evaluating 
the function associated to the initial perturbative factorially divergent power series explicitly via numerical methods. See for instance~\cite{Aniceto:2020ims} for an explicit application of this method. 
The Lambert-$W$ function 
previously appeared in the trans-asymptotic resummation of the similar Yukawa theory renormalon~\cite{BD} and in many other problems that are associated to renormalization and Dyson--Schwinger equations~\cite{borinsky2020euler,vanBaalen:2008tc,Klaczynski:2013fca,Panzer:2018tvy,Sberveglieri:2020eko}. 
Its ubiquitous appearance 
in Dyson--Schwinger type contexts deserves further explanation which lies beyond the scope of this article.

Our analysis is limited inherently by the set of diagrams that are captured by the Dyson--Schwinger equation~(\ref{dse}). For instance, it would be desirable to repeat it for a renormalon which also includes \emph{vertex-corrections}. A promising route out of this limitation is a combinatorial framework of Dyson--Schwinger equations~\cite{Kreimer:2006ua,foissy2008faa,vanBaalen:2008tc,vanBaalen:2009hu,Marie:2012cc,Kruger:2014eez,Kruger:2019tas,Bellon:2020uzi,Bellon:2020qlx} which allows for larger sets of diagrams to be included and which recently has been shown to allow for the extraction of asymptotic large-order information~\cite{Courtiel:2019dnq}. 

An illuminating comparison of instanton and renormalon phenomena based on available data in $\phi^4$ theory was recently put forward in~\cite{Dunne:2021lie}.
As there exists data on the large-order behaviour of $\phi^3$ theory in the minimal subtraction scheme~\cite{Mckane:1978me,McKane:2018ocs}, a similar study that compares renormalon effects, as studied in this article, 
and instanton effects in $\phi^3$ theory would be another feasible and interesting future endeavour. 

\subsection*{Acknowledgements} We thank Gerald Dunne, Dirk Kreimer and Max Meynig, for joint work that set up this problem,
Gerald Dunne for valuable comments on an early version of this article
and the Isaac Newton Institute in Cambridge, for remotely hosting the programme
{\em Applicable Resurgent Asymptotics} (ARA) that encouraged us to undertake this work. %
Moreover, we thank the anonymous referee 
for valuable comments that caused significant improvements of the manuscript.
    MB was supported by the NWO Vidi grant 680-47-551  ``Decoding Singularities of Feynman graphs'', Dr.\ Max R\"ossler, the Walter Haefner Foundation and the ETH Z\"urich Foundation.

\providecommand{\href}[2]{#2}\begingroup\raggedright\endgroup

\begin{thebibliography}{100}

\bibitem{Dyson:1952tj}
F.J.~Dyson, \emph{{Divergence of perturbation theory in quantum
  electrodynamics}}, \href{https://doi.org/10.1103/PhysRev.85.631}{\emph{Phys.
  Rev.} {\bfseries 85} (1952) 631}.

\bibitem{Lam:1968tk}
C.S.~Lam, \emph{{Behavior of very high order perturbation diagrams}},
  \href{https://doi.org/10.1007/BF02759226}{\emph{Nuovo Cim. A} {\bfseries 55}
  (1968) 258}.

\bibitem{Bender:1973rz}
C.M.~Bender and T.T.~Wu, \emph{Anharmonic oscillator. 2: A study of
  perturbation theory in large order},
  \href{https://doi.org/10.1103/PhysRevD.7.1620}{\emph{Phys. Rev. D} {\bfseries
  7} (1973) 1620}.

\bibitem{Lipatov:1976ny}
L.N.~Lipatov, \emph{Divergence of the perturbation theory series and the
  quasiclassical theory}, {\emph{Sov. Phys. JETP} {\bfseries 45} (1977) 216}.

\bibitem{Zinn-Justin:2004vcw}
J.~Zinn-Justin and U.D.~Jentschura, \emph{{Multi-instantons and exact results
  I: Conjectures, WKB expansions, and instanton interactions}},
  \href{https://doi.org/10.1016/j.aop.2004.04.004}{\emph{Annals Phys.}
  {\bfseries 313} (2004) 197}
  [\href{https://arxiv.org/abs/quant-ph/0501136}{{\ttfamily
  quant-ph/0501136}}].

\bibitem{Zinn-Justin:2004qzw}
J.~Zinn-Justin and U.D.~Jentschura, \emph{{Multi-instantons and exact results
  II: Specific cases, higher-order effects, and numerical calculations}},
  \href{https://doi.org/10.1016/j.aop.2004.04.003}{\emph{Annals Phys.}
  {\bfseries 313} (2004) 269}
  [\href{https://arxiv.org/abs/quant-ph/0501137}{{\ttfamily
  quant-ph/0501137}}].

\bibitem{Borinsky:2017hkb}
M.~Borinsky, \emph{{Renormalized asymptotic enumeration of Feynman diagrams}},
  \href{https://doi.org/10.1016/j.aop.2017.07.009}{\emph{Annals Phys.}
  {\bfseries 385} (2017) 95}
  [\href{https://arxiv.org/abs/1703.00840}{{\ttfamily 1703.00840}}].

\bibitem{borinsky2020euler}
M.~Borinsky and K.~Vogtmann, \emph{The {Euler} characteristic of
  {$\operatorname{Out}(F_n)$}},
  \href{https://doi.org/10.4171/CMH/501}{\emph{Commentarii Mathematici
  Helvetici} {\bfseries 95} (2020) 703}
  [\href{https://arxiv.org/abs/1907.03543}{{\ttfamily 1907.03543}}].

\bibitem{Zinn-Justin:1980oco}
J.~Zinn-Justin, \emph{Perturbation series at large orders in quantum mechanics
  and field theories: Application to the problem of resummation},
  \href{https://doi.org/10.1016/0370-1573(81)90016-8}{\emph{Phys. Rept.}
  {\bfseries 70} (1981) 109}.

\bibitem{le2012large}
J.-C.~Le~Guillou and J.~Zinn-Justin, \emph{Large-order behaviour of
  perturbation theory}, Elsevier (2012).

\bibitem{marino2015instantons}
M.~Mari\~no, \emph{Instantons and large N: an introduction to non-perturbative
  methods in quantum field theory}, Cambridge University Press (2015).

\bibitem{tHooft:1977xjm}
G.~'t~Hooft, \emph{Can we make sense out of quantum chromodynamics?},
  {\emph{Subnucl. Ser.} {\bfseries 15} (1979) 943}.

\bibitem{Lautrup:1977hs}
B.E.~Lautrup, \emph{On high order estimates in {QED}},
  \href{https://doi.org/10.1016/0370-2693(77)90145-9}{\emph{Phys. Lett. B}
  {\bfseries 69} (1977) 109}.

\bibitem{Parisi:1978bj}
G.~Parisi, \emph{Singularities of the {Borel} transform in renormalizable
  theories}, \href{https://doi.org/10.1016/0370-2693(78)90101-6}{\emph{Phys.
  Lett. B} {\bfseries 76} (1978) 65}.

\bibitem{Beneke:1998ui}
M.~Beneke, \emph{{Renormalons}},
  \href{https://doi.org/10.1016/S0370-1573(98)00130-6}{\emph{Phys. Rept.}
  {\bfseries 317} (1999) 1}
  [\href{https://arxiv.org/abs/hep-ph/9807443}{{\ttfamily hep-ph/9807443}}].

\bibitem{Shifman:2013uka}
M.~Shifman, \emph{New and old about renormalons: in memoriam {Kolya Uraltsev}},
  \href{https://doi.org/10.1142/S0217751X15430010}{\emph{Int. J. Mod. Phys. A}
  {\bfseries 30} (2015) 1543001}
  [\href{https://arxiv.org/abs/1310.1966}{{\ttfamily 1310.1966}}].

\bibitem{Palanques-Mestre:1983ogz}
A.~Palanques-Mestre and P.~Pascual, \emph{The {$1/N_f$} expansion of the
  $\gamma$ and $\beta$ functions in {QED}},
  \href{https://doi.org/10.1007/BF01212398}{\emph{Commun. Math. Phys.}
  {\bfseries 95} (1984) 277}.

\bibitem{Broadhurst:1992si}
D.J.~Broadhurst, \emph{{Large N expansion of QED: Asymptotic photon propagator
  and contributions to the muon anomaly, for any number of loops}},
  \href{https://doi.org/10.1007/BF01560355}{\emph{Z. Phys. C} {\bfseries 58}
  (1993) 339}.

\bibitem{Beneke:1992ea}
M.~Beneke and V.I.~Zakharov, \emph{{Improving large order perturbative
  expansions in quantum chromodynamics}},
  \href{https://doi.org/10.1103/PhysRevLett.69.2472}{\emph{Phys. Rev. Lett.}
  {\bfseries 69} (1992) 2472}.

\bibitem{Beneke:1993yn}
M.~Beneke and V.I.~Zakharov, \emph{{The first infrared renormalon in QED}},
  \href{https://doi.org/10.1016/0370-2693(93)91090-A}{\emph{Phys. Lett. B}
  {\bfseries 312} (1993) 340}.

\bibitem{Gracey:1996he}
J.A.~Gracey, \emph{{The QCD beta function at $O(1/N_f)$}},
  \href{https://doi.org/10.1016/0370-2693(96)00105-0}{\emph{Phys. Lett. B}
  {\bfseries 373} (1996) 178}
  [\href{https://arxiv.org/abs/hep-ph/9602214}{{\ttfamily hep-ph/9602214}}].

\bibitem{Dondi:2020qfj}
N.A.~Dondi, G.V.~Dunne, M.~Reichert and F.~Sannino, \emph{{Towards the QED beta
  function and renormalons at $1/N_f^2$ and $1/N_f^3$}},
  \href{https://doi.org/10.1103/PhysRevD.102.035005}{\emph{Phys. Rev. D}
  {\bfseries 102} (2020) 035005}
  [\href{https://arxiv.org/abs/2003.08397}{{\ttfamily 2003.08397}}].

\bibitem{Dondi:2021buw}
N.A.~Dondi, I.~Kalogerakis, D.~Orlando and S.~Reffert, \emph{{Resurgence of the
  large-charge expansion}},
  \href{https://doi.org/10.1007/JHEP05(2021)035}{\emph{JHEP} {\bfseries 05}
  (2021) 035} [\href{https://arxiv.org/abs/2102.12488}{{\ttfamily
  2102.12488}}].

\bibitem{DiPietro:2021yxb}
L.~Di~Pietro, M.~Mari\~no, G.~Sberveglieri and M.~Serone, \emph{Resurgence and
  {$1/N$} expansion in integrable field theories},
  \href{https://doi.org/10.1007/JHEP10(2021)166}{\emph{JHEP} {\bfseries 10}
  (2021) 166} [\href{https://arxiv.org/abs/2108.02647}{{\ttfamily
  2108.02647}}].

\bibitem{Fujimori:2021oqg}
T.~Fujimori, M.~Honda, S.~Kamata, T.~Misumi, N.~Sakai and T.~Yoda,
  \emph{{Quantum phase transition and resurgence: Lessons from
  three-dimensional $\mathcal{N}=4$ supersymmetric quantum electrodynamics}},
  \href{https://doi.org/10.1093/ptep/ptab086}{\emph{PTEP} {\bfseries 2021}
  (2021) 103B04} [\href{https://arxiv.org/abs/2103.13654}{{\ttfamily
  2103.13654}}].

\bibitem{Marino:2019eym}
M.~Mari\~no and T.~Reis, \emph{{Renormalons in integrable field theories}},
  \href{https://doi.org/10.1007/JHEP04(2020)160}{\emph{JHEP} {\bfseries 04}
  (2020) 160} [\href{https://arxiv.org/abs/1909.12134}{{\ttfamily
  1909.12134}}].

\bibitem{Marino:2019fvu}
M.~Mari\~no and T.~Reis, \emph{{A new renormalon in two dimensions}},
  \href{https://doi.org/10.1007/JHEP07(2020)216}{\emph{JHEP} {\bfseries 07}
  (2020) 216} [\href{https://arxiv.org/abs/1912.06228}{{\ttfamily
  1912.06228}}].

\bibitem{Marino:2021dzn}
M.~Mari\~no, R.~Miravitllas and T.~Reis, \emph{{New renormalons from analytic
  trans-series}},  \href{https://arxiv.org/abs/2111.11951}{{\ttfamily
  2111.11951}}.

\bibitem{Maiezza:2018pkk}
A.~Maiezza and J.C.~Vasquez, \emph{Renormalons in a general quantum field
  theory}, \href{https://doi.org/10.1016/j.aop.2018.04.027}{\emph{Annals Phys.}
  {\bfseries 394} (2018) 84}
  [\href{https://arxiv.org/abs/1802.06022}{{\ttfamily 1802.06022}}].

\bibitem{Antipin:2018asc}
O.~Antipin, A.~Maiezza and J.C.~Vasquez, \emph{{Resummation in QFT with Meijer
  G-functions}},
  \href{https://doi.org/10.1016/j.nuclphysb.2019.02.014}{\emph{Nucl. Phys. B}
  {\bfseries 941} (2019) 72}
  [\href{https://arxiv.org/abs/1807.05060}{{\ttfamily 1807.05060}}].

\bibitem{Maiezza:2020nbe}
A.~Maiezza and J.C.~Vasquez, \emph{{Non-Wilsonian ultraviolet completion via
  transseries}}, \href{https://doi.org/10.1142/S0217751X21500160}{\emph{Int. J.
  Mod. Phys. A} {\bfseries 36} (2021) 2150016}
  [\href{https://arxiv.org/abs/2007.01270}{{\ttfamily 2007.01270}}].

\bibitem{Maiezza:2021mry}
A.~Maiezza and J.C.~Vasquez, \emph{{Resurgence of the QCD Adler function}},
  \href{https://doi.org/10.1016/j.physletb.2021.136338}{\emph{Phys. Lett. B}
  {\bfseries 817} (2021) 136338}
  [\href{https://arxiv.org/abs/2104.03095}{{\ttfamily 2104.03095}}].

\bibitem{Caprini:2020lff}
I.~Caprini, \emph{{Conformal mapping of the Borel plane: going beyond
  perturbative QCD}},
  \href{https://doi.org/10.1103/PhysRevD.102.054017}{\emph{Phys. Rev. D}
  {\bfseries 102} (2020) 054017}
  [\href{https://arxiv.org/abs/2006.16605}{{\ttfamily 2006.16605}}].

\bibitem{Pazarbasi:2019web}
C.~Pazarba\c{s}\i{} and D.~Van Den~Bleeken, \emph{{Renormalons in quantum
  mechanics}}, \href{https://doi.org/10.1007/JHEP08(2019)096}{\emph{JHEP}
  {\bfseries 08} (2019) 096}
  [\href{https://arxiv.org/abs/1906.07198}{{\ttfamily 1906.07198}}].

\bibitem{Cavalcanti:2020osb}
E.~Cavalcanti, \emph{{Renormalons beyond the Borel plane}},
  \href{https://doi.org/10.1103/PhysRevD.103.025019}{\emph{Phys. Rev. D}
  {\bfseries 103} (2021) 025019}
  [\href{https://arxiv.org/abs/2011.11175}{{\ttfamily 2011.11175}}].

\bibitem{Balduf:2021kag}
P.-H.~Balduf, \emph{{Dyson--Schwinger equations in minimal subtraction}},
  \href{https://arxiv.org/abs/2109.13684}{{\ttfamily 2109.13684}}.

\bibitem{Macfarlane:1974vp}
A.J.~Macfarlane and G.~Woo, \emph{$\phi^3$ theory in six dimensions and the
  renormalization group},
  \href{https://doi.org/10.1016/0550-3213(74)90306-X}{\emph{Nucl. Phys. B}
  {\bfseries 77} (1974) 91}.

\bibitem{Gracey:2015tta}
J.A.~Gracey, \emph{{Four loop renormalization of $\phi^3$ theory in six
  dimensions}}, \href{https://doi.org/10.1103/PhysRevD.92.025012}{\emph{Phys.
  Rev. D} {\bfseries 92} (2015) 025012}
  [\href{https://arxiv.org/abs/1506.03357}{{\ttfamily 1506.03357}}].

\bibitem{Borinsky:2021jdb}
M.~Borinsky, J.A.~Gracey, M.V.~Kompaniets and O.~Schnetz, \emph{{Five-loop
  renormalization of $\phi^3$ theory with applications to the Lee--Yang edge
  singularity and percolation theory}},
  \href{https://doi.org/10.1103/PhysRevD.103.116024}{\emph{Phys. Rev. D}
  {\bfseries 103} (2021) 116024}
  [\href{https://arxiv.org/abs/2103.16224}{{\ttfamily 2103.16224}}].

\bibitem{Borinsky:2021gkd}
M.~Borinsky and O.~Schnetz, \emph{{Graphical functions in even dimensions}},
  {\emph{Commun. Number Theory Phys.} (2022) to appear}
  [\href{https://arxiv.org/abs/2105.05015}{{\ttfamily 2105.05015}}].

\bibitem{Kompaniets:2021hwg}
M.~Kompaniets and A.~Pikelner, \emph{{Critical exponents from five-loop scalar
  theory renormalization near six-dimensions}},
  \href{https://doi.org/10.1016/j.physletb.2021.136331}{\emph{Phys. Lett. B}
  {\bfseries 817} (2021) 136331}
  [\href{https://arxiv.org/abs/2101.10018}{{\ttfamily 2101.10018}}].

\bibitem{Borinsky:2022}
M.~Borinsky and O.~Schnetz, \emph{Recursive computation of {Feynman} periods},
  {\emph{in preparation} (2022) }.

\bibitem{Brezin:1976vw}
E.~Brezin, J.C.~Le~Guillou and J.~Zinn-Justin, \emph{Perturbation theory at
  large order. 1. {The} $\phi^{2N}$ interaction},
  \href{https://doi.org/10.1103/PhysRevD.15.1544}{\emph{Phys. Rev. D}
  {\bfseries 15} (1977) 1544}.

\bibitem{Mckane:1978me}
A.J.~McKane, \emph{Vacuum instability in scalar field theories},
  \href{https://doi.org/10.1016/0550-3213(79)90086-5}{\emph{Nucl. Phys. B}
  {\bfseries 152} (1979) 166}.

\bibitem{Houghton:1978dt}
A.~Houghton, J.S.~Reeve and D.J.~Wallace, \emph{High order behavior in $\phi^3$
  field theories and the percolation problem},
  \href{https://doi.org/10.1103/PhysRevB.17.2956}{\emph{Phys. Rev. B}
  {\bfseries 17} (1978) 2956}.

\bibitem{alvarez1988coupling}
G.~Álvarez, \emph{Coupling-constant behavior of the resonances of the cubic
  anharmonic oscillator},
  \href{https://doi.org/10.1103/PhysRevA.37.4079}{\emph{Physical Review A}
  {\bfseries 37} (1988) 4079}.

\bibitem{alvarez1995bender}
G.~Álvarez, \emph{{Bender--Wu} branch points in the cubic oscillator},
  \href{https://doi.org/10.1088/0305-4470/28/16/016}{\emph{Journal of Physics
  A: Mathematical and General} {\bfseries 28} (1995) 4589}.

\bibitem{Bern:2008qj}
Z.~Bern, J.J.M.~Carrasco and H.~Johansson, \emph{New relations for gauge-theory
  amplitudes}, \href{https://doi.org/10.1103/PhysRevD.78.085011}{\emph{Phys.
  Rev. D} {\bfseries 78} (2008) 085011}
  [\href{https://arxiv.org/abs/0805.3993}{{\ttfamily 0805.3993}}].

\bibitem{delaCruz:2017zqr}
L.~de~la Cruz, A.~Kniss and S.~Weinzierl, \emph{{Properties of scattering forms
  and their relation to associahedra}},
  \href{https://doi.org/10.1007/JHEP03(2018)064}{\emph{JHEP} {\bfseries 03}
  (2018) 064} [\href{https://arxiv.org/abs/1711.07942}{{\ttfamily
  1711.07942}}].

\bibitem{Arkani-Hamed:2017mur}
N.~Arkani-Hamed, Y.~Bai, S.~He and G.~Yan, \emph{Scattering forms and the
  positive geometry of kinematics, color and the worldsheet},
  \href{https://doi.org/10.1007/JHEP05(2018)096}{\emph{JHEP} {\bfseries 05}
  (2018) 096} [\href{https://arxiv.org/abs/1711.09102}{{\ttfamily
  1711.09102}}].

\bibitem{Bonfim_1980}
O.F.~de~Alcantara~Bonfim, J.E.~Kirkham and A.J.~McKane, \emph{Critical
  exponents to order $\epsilon^3$ for $\phi^3$ models of critical phenomena in
  $6-\epsilon$ dimensions},
  \href{https://doi.org/10.1088/0305-4470/13/7/006}{\emph{J. Phys. A}
  {\bfseries 13} (1980) L247}.

\bibitem{Bonfim_1981}
O.F.~de~Alcantara~Bonfim, J.E.~Kirkham and A.J.~McKane, \emph{Critical
  exponents for the percolation problem and the {Yang--Lee} edge singularity},
  \href{https://doi.org/10.1088/0305-4470/14/9/034}{\emph{J. Phys. A}
  {\bfseries 14} (1981) 2391}.

\bibitem{Fisher_1978}
M.E.~Fisher, \emph{{Yang--Lee} edge singularity and ${\ensuremath{\phi}}^{3}$
  field theory}, \href{https://doi.org/10.1103/PhysRevLett.40.1610}{\emph{Phys.
  Rev. Lett.} {\bfseries 40} (1978) 1610}.

\bibitem{Connes:1999yr}
A.~Connes and D.~Kreimer, \emph{{Renormalization in quantum field theory and
  the Riemann-Hilbert problem. 1. The Hopf algebra structure of graphs and the
  main theorem}}, \href{https://doi.org/10.1007/s002200050779}{\emph{Commun.
  Math. Phys.} {\bfseries 210} (2000) 249}
  [\href{https://arxiv.org/abs/hep-th/9912092}{{\ttfamily hep-th/9912092}}].

\bibitem{BK2}
D.~Broadhurst and D.~Kreimer, \emph{{Exact solutions of Dyson--Schwinger
  equations for iterated one loop integrals and propagator coupling duality}},
  \href{https://doi.org/10.1016/S0550-3213(01)00071-2}{\emph{Nucl. Phys. B}
  {\bfseries 600} (2001) 403}
  [\href{https://arxiv.org/abs/hep-th/0012146}{{\ttfamily hep-th/0012146}}].

\bibitem{BDM}
M.~Borinsky, G.V.~Dunne and M.~Meynig, \emph{{Semiclassical trans-series from
  the perturbative Hopf-algebraic Dyson--Schwinger equations: $\phi^3$ QFT in 6
  dimensions}}, \href{https://doi.org/10.3842/SIGMA.2021.087}{\emph{SIGMA}
  {\bfseries 17} (2021) 087}
  [\href{https://arxiv.org/abs/2104.00593}{{\ttfamily 2104.00593}}].

\bibitem{ecalle1981fonctions}
J.~{\'E}calle, \emph{Les fonctions r{\'e}surgentes}, vol.~1, Universit{\'e} de
  Paris-Sud, D{\'e}partement de Math{\'e}matique, B{\^a}t. 425 (1981).

\bibitem{Mitschi:2016fxp}
C.~Mitschi and D.~Sauzin, \emph{{Divergent Series, Summability and Resurgence
  I}}, vol.~2153, Springer (2016),
  \href{https://doi.org/10.1007/978-3-319-28736-2}{10.1007/978-3-319-28736-2}.

\bibitem{Marino:2012zq}
M.~Mari\~no, \emph{Lectures on non-perturbative effects in large {$N$} gauge
  theories, matrix models and strings},
  \href{https://doi.org/10.1002/prop.201400005}{\emph{Fortsch. Phys.}
  {\bfseries 62} (2014) 455} [\href{https://arxiv.org/abs/1206.6272}{{\ttfamily
  1206.6272}}].

\bibitem{Dorigoni:2014hea}
D.~Dorigoni, \emph{An introduction to resurgence, trans-series and alien
  calculus}, \href{https://doi.org/10.1016/j.aop.2019.167914}{\emph{Annals
  Phys.} {\bfseries 409} (2019) 167914}
  [\href{https://arxiv.org/abs/1411.3585}{{\ttfamily 1411.3585}}].

\bibitem{Aniceto:2018bis}
I.~Aniceto, G.~Basar and R.~Schiappa, \emph{A primer on resurgent transseries
  and their asymptotics},
  \href{https://doi.org/10.1016/j.physrep.2019.02.003}{\emph{Phys. Rept.}
  {\bfseries 809} (2019) 1} [\href{https://arxiv.org/abs/1802.10441}{{\ttfamily
  1802.10441}}].

\bibitem{Delabaere1997}
E.~Delabaere, H.~Dillinger and F.~Pham, \emph{Exact semiclassical expansions
  for one-dimensional quantum oscillators},
  \href{https://doi.org/10.1063/1.532206}{\emph{Journal of Mathematical
  Physics} {\bfseries 38} (1997) 6126}.

\bibitem{delabaere99}
E.~Delabaere and F.~Pham, \emph{Resurgent methods in semi-classical
  asymptotics}, {\emph{Annales de l'I.H.P. Physique th\'eorique} {\bfseries 71}
  (1999) 1}.

\bibitem{alvarez04}
G.~Álvarez, \emph{{Langer–Cherry} derivation of the multi-instanton
  expansion for the symmetric double well},
  \href{https://doi.org/10.1063/1.1767988}{\emph{Journal of Mathematical
  Physics} {\bfseries 45} (2004) 3095}.

\bibitem{Argyres:2012ka}
P.C.~Argyres and M.~\"Unsal, \emph{{The semi-classical expansion and resurgence
  in gauge theories: new perturbative, instanton, bion, and renormalon
  effects}}, \href{https://doi.org/10.1007/JHEP08(2012)063}{\emph{JHEP}
  {\bfseries 08} (2012) 063} [\href{https://arxiv.org/abs/1206.1890}{{\ttfamily
  1206.1890}}].

\bibitem{Dunne:2012ae}
G.V.~Dunne and M.~\"Unsal, \emph{Resurgence and trans-series in quantum field
  theory: The $\mathbb{CP}^{N-1}$ model},
  \href{https://doi.org/10.1007/JHEP11(2012)170}{\emph{JHEP} {\bfseries 11}
  (2012) 170} [\href{https://arxiv.org/abs/1210.2423}{{\ttfamily 1210.2423}}].

\bibitem{Dunne:2013ada}
G.V.~Dunne and M.~\"Unsal, \emph{{Generating nonperturbative physics from
  perturbation theory}},
  \href{https://doi.org/10.1103/PhysRevD.89.041701}{\emph{Phys. Rev. D}
  {\bfseries 89} (2014) 041701}
  [\href{https://arxiv.org/abs/1306.4405}{{\ttfamily 1306.4405}}].

\bibitem{Marino:2007te}
M.~Mari\~no, R.~Schiappa and M.~Weiss, \emph{Nonperturbative effects and the
  large-order behavior of matrix models and topological strings},
  \href{https://doi.org/10.4310/CNTP.2008.v2.n2.a3}{\emph{Commun. Num. Theor.
  Phys.} {\bfseries 2} (2008) 349}
  [\href{https://arxiv.org/abs/0711.1954}{{\ttfamily 0711.1954}}].

\bibitem{Pasquetti:2010bps}
S.~Pasquetti and R.~Schiappa, \emph{{Borel and Stokes} nonperturbative
  phenomena in topological string theory and $c=1$ matrix models},
  \href{https://doi.org/10.1007/s00023-010-0044-5}{\emph{Annales Henri
  Poincare} {\bfseries 11} (2010) 351}
  [\href{https://arxiv.org/abs/0907.4082}{{\ttfamily 0907.4082}}].

\bibitem{Aniceto:2011nu}
I.~Aniceto, R.~Schiappa and M.~Vonk, \emph{The resurgence of instantons in
  string theory},
  \href{https://doi.org/10.4310/CNTP.2012.v6.n2.a3}{\emph{Commun. Num. Theor.
  Phys.} {\bfseries 6} (2012) 339}
  [\href{https://arxiv.org/abs/1106.5922}{{\ttfamily 1106.5922}}].

\bibitem{Grassi:2014cla}
A.~Grassi, M.~Mari\~no and S.~Zakany, \emph{{Resumming the string perturbation
  series}}, \href{https://doi.org/10.1007/JHEP05(2015)038}{\emph{JHEP}
  {\bfseries 05} (2015) 038} [\href{https://arxiv.org/abs/1405.4214}{{\ttfamily
  1405.4214}}].

\bibitem{Marino:2002fk}
M.~Mari\~no, \emph{{Chern-Simons theory, matrix integrals, and perturbative
  three manifold invariants}},
  \href{https://doi.org/10.1007/s00220-004-1194-4}{\emph{Commun. Math. Phys.}
  {\bfseries 253} (2004) 25}
  [\href{https://arxiv.org/abs/hep-th/0207096}{{\ttfamily hep-th/0207096}}].

\bibitem{Andersen:2018khh}
J.E.~Andersen and W.E.~Misteg\r{a}rd, \emph{Resurgence analysis of quantum
  invariants of {Seifert} fibered homology spheres},
  \href{https://arxiv.org/abs/1811.05376}{{\ttfamily 1811.05376}}.

\bibitem{Eynard:2019mps}
B.~Eynard, \emph{{Large genus behavior of topological recursion}},
  \href{https://arxiv.org/abs/1905.11270}{{\ttfamily 1905.11270}}.

\bibitem{Kontsevich:2020piu}
M.~Kontsevich and Y.~Soibelman, \emph{{Analyticity and resurgence in
  wall-crossing formulas}},  \href{https://arxiv.org/abs/2005.10651}{{\ttfamily
  2005.10651}}.

\bibitem{Garoufalidis:2020nut}
S.~Garoufalidis, J.~Gu and M.~Mari\~no, \emph{The resurgent structure of
  quantum knot invariants},
  \href{https://doi.org/10.1007/s00220-021-04076-0}{\emph{Commun. Math. Phys.}
  {\bfseries 386} (2021) 469}
  [\href{https://arxiv.org/abs/2007.10190}{{\ttfamily 2007.10190}}].

\bibitem{borinsky2018generating}
M.~Borinsky, \emph{{Generating asymptotics for factorially divergent
  sequences}}, \href{https://doi.org/10.37236/5999}{\emph{The Electronic
  Journal of Combinatorics} {\bfseries 25} (2018) 4}
  [\href{https://arxiv.org/abs/1603.01236}{{\ttfamily 1603.01236}}].

\bibitem{Bellon:2014zxa}
M.P.~Bellon and P.J.~Clavier, \emph{A {Schwinger--Dyson} equation in the
  {Borel} plane: Singularities of the solution},
  \href{https://doi.org/10.1007/s11005-015-0761-2}{\emph{Lett. Math. Phys.}
  {\bfseries 105} (2015) 795}
  [\href{https://arxiv.org/abs/1411.7190}{{\ttfamily 1411.7190}}].

\bibitem{Bellon:2016mje}
M.P.~Bellon and P.J.~Clavier, \emph{Alien calculus and a {Schwinger--Dyson}
  equation: two-point function with a nonperturbative mass scale},
  \href{https://doi.org/10.1007/s11005-017-1016-1}{\emph{Lett. Math. Phys.}
  {\bfseries 108} (2018) 391}
  [\href{https://arxiv.org/abs/1612.07813}{{\ttfamily 1612.07813}}].

\bibitem{Clavier:2019sph}
P.J.~Clavier, \emph{{Borel-\'Ecalle} resummation of a two-point function},
  \href{https://doi.org/10.1007/s00023-021-01057-w}{\emph{Annales Henri
  Poincare} {\bfseries 22} (2021) 2103}
  [\href{https://arxiv.org/abs/1912.03237}{{\ttfamily 1912.03237}}].

\bibitem{Bellon:2020uzi}
M.P.~Bellon and E.I.~Russo, \emph{Resurgent analysis of
  {Ward--Schwinger--Dyson} equations},
  \href{https://doi.org/10.3842/SIGMA.2021.075}{\emph{SIGMA} {\bfseries 17}
  (2021) 075} [\href{https://arxiv.org/abs/2011.13822}{{\ttfamily
  2011.13822}}].

\bibitem{Bellon:2020qlx}
M.P.~Bellon and E.I.~Russo, \emph{{Ward--Schwinger--Dyson} equations in
  $\phi^3_6$ quantum field theory},
  \href{https://doi.org/10.1007/s11005-021-01377-2}{\emph{Lett. Math. Phys.}
  {\bfseries 111} (2021) 42}
  [\href{https://arxiv.org/abs/2007.15675}{{\ttfamily 2007.15675}}].

\bibitem{BD}
M.~Borinsky and G.V.~Dunne, \emph{Non-perturbative completion of
  {Hopf}-algebraic {Dyson--Schwinger} equations},
  \href{https://doi.org/10.1016/j.nuclphysb.2020.115096}{\emph{Nucl. Phys. B}
  {\bfseries 957} (2020) 115096}
  [\href{https://arxiv.org/abs/2005.04265}{{\ttfamily 2005.04265}}].

\bibitem{BK1}
D.~Broadhurst and D.~Kreimer, \emph{{Combinatoric explosion of renormalization
  tamed by Hopf algebra: Thirty loop Pade--Borel resummation}},
  \href{https://doi.org/10.1016/S0370-2693(00)00051-4}{\emph{Phys. Lett. B}
  {\bfseries 475} (2000) 63}
  [\href{https://arxiv.org/abs/hep-th/9912093}{{\ttfamily hep-th/9912093}}].

\bibitem{PainI}
S.~Garoufalidis, A.~Its, A.~Kapaev and M.~Mari\~no, \emph{{Asymptotics of the
  instantons of Painlev\'e I}},
  \href{https://doi.org/10.1093/imrn/rnr029}{\emph{Int. Math. Res. Not.}
  {\bfseries 2012} (2012) 561}
  [\href{https://arxiv.org/abs/1002.3634}{{\ttfamily 1002.3634}}].

\bibitem{Pain1a}
I.~Aniceto, R.~Schiappa and M.~Vonk, \emph{The resurgence of instantons in
  string theory},
  \href{https://doi.org/10.4310/CNTP.2012.v6.n2.a3}{\emph{Commun. Num. Theor.
  Phys.} {\bfseries 6} (2012) 339}
  [\href{https://arxiv.org/abs/1106.5922}{{\ttfamily 1106.5922}}].

\bibitem{Marino:2006hs}
M.~Mari\~no, \emph{{Open string amplitudes and large order behavior in
  topological string theory}},
  \href{https://doi.org/10.1088/1126-6708/2008/03/060}{\emph{JHEP} {\bfseries
  03} (2008) 060} [\href{https://arxiv.org/abs/hep-th/0612127}{{\ttfamily
  hep-th/0612127}}].

\bibitem{PhysRevA.33.12}
J.c.v.~C\'{\i}\ifmmode~\check{z}\else \v{z}\fi{}ek, R.J.~Damburg, S.~Graffi,
  V.~Grecchi, E.M.~Harrell, J.G.~Harris et~al., \emph{{$1/R$} expansion for
  {${\mathrm{H}}_{2}^{+}$}: Calculation of exponentially small terms and
  asymptotics}, \href{https://doi.org/10.1103/PhysRevA.33.12}{\emph{Phys. Rev.
  A} {\bfseries 33} (1986) 12}.

\bibitem{Zinn-Justin:1979jnt}
J.~Zinn-Justin, \emph{Expansion around instantons in quantum mechanics},
  \href{https://doi.org/10.1063/1.524919}{\emph{J. Math. Phys.} {\bfseries 22}
  (1981) 511}.

\bibitem{Pazarbasi:2021ifb}
C.~Pazarba\c{s}\i{} and M.~\"Unsal, \emph{Cluster expansion and resurgence in
  the {Polyakov} model},
  \href{https://doi.org/10.1103/PhysRevLett.128.151601}{\emph{Phys. Rev. Lett.}
  {\bfseries 128} (2022) 151601}
  [\href{https://arxiv.org/abs/2110.05612}{{\ttfamily 2110.05612}}].

\bibitem{Dunne:2016qix}
G.V.~Dunne and M.~\"Unsal, \emph{{WKB} and resurgence in the {Mathieu}
  equation},  in \emph{Resurgence, Physics and Numbers}, pp.~249--298, Springer
  (2017), \href{https://doi.org/10.1007/978-88-7642-613-1_6}{DOI}
  [\href{https://arxiv.org/abs/1603.04924}{{\ttfamily 1603.04924}}].

\bibitem{dingle1973asymptotic}
R.B.~Dingle, \emph{Asymptotic expansions: their derivation and interpretation},
  vol.~521, Academic Press London (1973).

\bibitem{berry1991hyperasymptotics}
M.V.~Berry and C.J.~Howls, \emph{Hyperasymptotics for integrals with saddles},
  {\emph{Proceedings of the Royal Society of London. Series A: Mathematical and
  Physical Sciences} {\bfseries 434} (1991) 657}.

\bibitem{costin2008asymptotics}
O.~Costin, \emph{Asymptotics and Borel summability}, CRC press (2008).

\bibitem{Aniceto:2013fka}
I.~Aniceto and R.~Schiappa, \emph{Nonperturbative ambiguities and the reality
  of resurgent transseries},
  \href{https://doi.org/10.1007/s00220-014-2165-z}{\emph{Commun. Math. Phys.}
  {\bfseries 335} (2015) 183}
  [\href{https://arxiv.org/abs/1308.1115}{{\ttfamily 1308.1115}}].

\bibitem{youtubeGerald}
G.~Dunne, \emph{{Resurgent trans-series analysis of Hopf algebraic
  renormalization}}, {\emph{Talk at IHES conference Algebraic Structures in
  Perturbative Quantum Field Theories} {\bfseries Video:
  \url{https://www.youtube.com/watch?v=29ENkKqUDvI#t=37m00s}} (19 Nov 2020) }.

\bibitem{Caliceti:2007ra}
E.~Caliceti, M.~Meyer-Hermann, P.~Ribeca, A.~Surzhykov and U.D.~Jentschura,
  \emph{{From useful algorithms for slowly convergent series to physical
  predictions based on divergent perturbative expansions}},
  \href{https://doi.org/10.1016/j.physrep.2007.03.003}{\emph{Phys. Rept.}
  {\bfseries 446} (2007) 1} [\href{https://arxiv.org/abs/0707.1596}{{\ttfamily
  0707.1596}}].

\bibitem{Costin:2020pcj}
O.~Costin and G.V.~Dunne, \emph{Uniformization and constructive analytic
  continuation of taylor series},
  \href{https://doi.org/10.1007/s00220-022-04361-6}{\emph{Commun. Math. Phys.}
  {\bfseries 392} (2022) 863}
  [\href{https://arxiv.org/abs/2009.01962}{{\ttfamily 2009.01962}}].

\bibitem{Costin:2020hwg}
O.~Costin and G.V.~Dunne, \emph{Physical resurgent extrapolation},
  \href{https://doi.org/10.1016/j.physletb.2020.135627}{\emph{Phys. Lett. B}
  {\bfseries 808} (2020) 135627}
  [\href{https://arxiv.org/abs/2003.07451}{{\ttfamily 2003.07451}}].

\bibitem{vanBaalen:2008tc}
G.~van Baalen, D.~Kreimer, D.~Uminsky and K.~Yeats, \emph{{The QED
  beta-function from global solutions to Dyson--Schwinger equations}},
  \href{https://doi.org/10.1016/j.aop.2008.05.007}{\emph{Annals Phys.}
  {\bfseries 324} (2009) 205}
  [\href{https://arxiv.org/abs/0805.0826}{{\ttfamily 0805.0826}}].

\bibitem{Klaczynski:2013fca}
L.~Klaczynski and D.~Kreimer, \emph{Avoidance of a {Landau} pole by flat
  contributions in {QED}},
  \href{https://doi.org/10.1016/j.aop.2014.02.019}{\emph{Annals Phys.}
  {\bfseries 344} (2014) 213}
  [\href{https://arxiv.org/abs/1309.5061}{{\ttfamily 1309.5061}}].

\bibitem{Panzer:2018tvy}
E.~Panzer and R.~Wulkenhaar, \emph{Lambert-{W} solves the noncommutative
  $\varphi ^4$-model},
  \href{https://doi.org/10.1007/s00220-019-03592-4}{\emph{Commun. Math. Phys.}
  {\bfseries 374} (2019) 1935}
  [\href{https://arxiv.org/abs/1807.02945}{{\ttfamily 1807.02945}}].

\bibitem{Sberveglieri:2020eko}
G.~Sberveglieri, M.~Serone and G.~Spada, \emph{Self-dualities and
  renormalization dependence of the phase diagram in 3d {$O(N)$} vector
  models}, \href{https://doi.org/10.1007/JHEP02(2021)098}{\emph{JHEP}
  {\bfseries 02} (2021) 098}
  [\href{https://arxiv.org/abs/2010.09737}{{\ttfamily 2010.09737}}].

\bibitem{borinsky2018graphs}
M.~Borinsky, \emph{Graphs in perturbation theory: Algebraic structure and
  asymptotics}, Springer (2018),
  \href{https://doi.org/10.1007/978-3-030-03541-9}{10.1007/978-3-030-03541-9}.

\bibitem{GP}
{The PARI Group, Univ. Bordeaux}, \emph{{PARI/GP version \texttt{2.13.1}}}.
\newblock \url{http://pari.math.u-bordeaux.fr/}, 2021.

\bibitem{Mahmoud:2020vww}
A.A.~Mahmoud and K.~Yeats, \emph{Connected chord diagrams and the combinatorics
  of asymptotic expansions},
  \href{https://arxiv.org/abs/2010.06550}{{\ttfamily 2010.06550}}.

\bibitem{mahmoud2020asymptotic}
A.A.~Mahmoud, \emph{An asymptotic expansion for the number of 2-connected chord
  diagrams},  \href{https://arxiv.org/abs/2009.12688}{{\ttfamily 2009.12688}}.

\bibitem{Mahmoud:2020qbb}
A.A.~Mahmoud, \emph{On Enumerative Structures in Quantum Field Theory}, Ph.D.
  thesis, U. Waterloo, 2020.
\newblock \href{https://arxiv.org/abs/2008.11661}{{\ttfamily 2008.11661}}.

\bibitem{AssemMahmoud:2020rvk}
A.A.~Mahmoud, \emph{New prospects of enumeration in quantum electrodynamics},
  \href{https://arxiv.org/abs/2011.04291}{{\ttfamily 2011.04291}}.

\bibitem{Aniceto:2020ims}
I.~Aniceto, D.~Hasenbichler, C.J.~Howls and C.J.~Lustri, \emph{{Capturing the
  cascade: a transseries approach to delayed bifurcations}},
  \href{https://doi.org/10.1088/1361-6544/ac2e44}{\emph{Nonlinearity}
  {\bfseries 34} (2021) 8248}
  [\href{https://arxiv.org/abs/2012.09779}{{\ttfamily 2012.09779}}].

\bibitem{Kreimer:2006ua}
D.~Kreimer and K.~Yeats, \emph{{An \'etude in non-linear Dyson--Schwinger
  equations}},
  \href{https://doi.org/10.1016/j.nuclphysbps.2006.09.036}{\emph{Nucl. Phys. B
  Proc. Suppl.} {\bfseries 160} (2006) 116}
  [\href{https://arxiv.org/abs/hep-th/0605096}{{\ttfamily hep-th/0605096}}].

\bibitem{foissy2008faa}
L.~Foissy, \emph{{Fa{\`a} di Bruno} subalgebras of the {Hopf} algebra of planar
  trees from combinatorial {Dyson--Schwinger} equations},
  \href{https://doi.org/10.1016/J.AIM.2007.12.003}{\emph{Advances in
  Mathematics} {\bfseries 218} (2008) 136}
  [\href{https://arxiv.org/abs/0707.1204}{{\ttfamily 0707.1204}}].

\bibitem{vanBaalen:2009hu}
G.~van Baalen, D.~Kreimer, D.~Uminsky and K.~Yeats, \emph{{The QCD
  beta-function from global solutions to Dyson--Schwinger equations}},
  \href{https://doi.org/10.1016/j.aop.2009.10.011}{\emph{Annals Phys.}
  {\bfseries 325} (2010) 300}
  [\href{https://arxiv.org/abs/0906.1754}{{\ttfamily 0906.1754}}].

\bibitem{Marie:2012cc}
N.~Marie and K.~Yeats, \emph{{A chord diagram expansion coming from some
  Dyson--Schwinger equations}},
  \href{https://doi.org/10.4310/CNTP.2013.v7.n2.a2}{\emph{Commun. Num. Theor
  Phys.} {\bfseries 07} (2013) 251}
  [\href{https://arxiv.org/abs/1210.5457}{{\ttfamily 1210.5457}}].

\bibitem{Kruger:2014eez}
O.~Kr\"uger and D.~Kreimer, \emph{Filtrations in {Dyson--Schwinger} equations:
  Next-to$^{j}$-leading log expansions systematically},
  \href{https://doi.org/10.1016/j.aop.2015.05.013}{\emph{Annals Phys.}
  {\bfseries 360} (2015) 293}
  [\href{https://arxiv.org/abs/1412.1657}{{\ttfamily 1412.1657}}].

\bibitem{Kruger:2019tas}
O.~Kr\"uger, \emph{{Log expansions from combinatorial Dyson--Schwinger
  equations}}, \href{https://doi.org/10.1007/s11005-020-01288-8}{\emph{Lett.
  Math. Phys.} {\bfseries 110} (2020) 2175}
  [\href{https://arxiv.org/abs/1906.06131}{{\ttfamily 1906.06131}}].

\bibitem{Courtiel:2019dnq}
J.~Courtiel and K.~Yeats, \emph{$\hbox {Next-to}{}^k$ leading log expansions by
  chord diagrams},
  \href{https://doi.org/10.1007/s00220-020-03691-7}{\emph{Commun. Math. Phys.}
  {\bfseries 377} (2020) 469}
  [\href{https://arxiv.org/abs/1906.05139}{{\ttfamily 1906.05139}}].

\bibitem{Dunne:2021lie}
G.V.~Dunne and M.~Meynig, \emph{{Instantons or renormalons? Remarks on
  $\phi^4_{d=4}$ theory in the MS scheme}},
  \href{https://doi.org/10.1103/PhysRevD.105.025019}{\emph{Phys. Rev. D}
  {\bfseries 105} (2022) 025019}
  [\href{https://arxiv.org/abs/2111.15554}{{\ttfamily 2111.15554}}].

\bibitem{McKane:2018ocs}
A.J.~McKane, \emph{{Perturbation expansions at large order: Results for scalar
  field theories revisited}},
  \href{https://doi.org/10.1088/1751-8121/aaf768}{\emph{J. Phys. A} {\bfseries
  52} (2019) 055401} [\href{https://arxiv.org/abs/1807.00656}{{\ttfamily
  1807.00656}}].

\end{thebibliography}
\end{document}